\documentclass[aps,prx,reprint]{revtex4-2}

\usepackage[colorlinks=true,citecolor=blue]{hyperref}
\usepackage{graphicx}
\usepackage{amsmath}
\usepackage{amsfonts}
\usepackage{comment}
\usepackage{verbatim}
\usepackage{mathtools}

\begin{document}

\title{Optimal Search Strategies under Finite Resource Constraints: The Role of Environmental Memory and Directional Information in Informed and Blind Protocols}

\author{Hugues Meyer}
\affiliation{School of Physics and Astronomy, University of Nottingham, Nottingham NG7 2RD, United Kingdom}
\affiliation{Centre for the Mathematics and Theoretical Physics of Quantum Non-Equilibrium Systems, University of Nottingham, Nottingham NG7 2RD, United Kingdom}
\begin{abstract}
Improving the efficiency of a stochastic search often requires to provide the searching agent with resources that modify its dynamics and help it explore space more quickly. Although such resources are often assumed to be unlimited and treated as a cost to mitigate, many realistic search processes operate under a finite resource budget, e.g. in processes with environmental memory in which the searcher deposits long-lasting cues along its path that it can interact with at later times to guide its further search.  Here, we address the question of how to allocate a limited budget to maximize search efficiency by investigating a one-dimensional diffusing agent whose access to resources is constrained. We consider three resource models, namely an external force applied by an external controller to pull the searcher towards the target,  directional flags deposited by the searcher along its path that redirect its motion towards the target, and self-generated chemorepulsive cues that act as a force field. For all these models, we find that resource usage should be concentrated in narrow regions of space when information about the searcher's current and past positions is available. In contrast, in blind strategies where a predefined time protocol is followed, resources should be used at a steady rate to ensure it to be used at most of its capacity by the time it reaches the target. We further show that blind strategies in non-Markovian processes with environmental memory are significantly less efficient due to the inherent fluctuations in the locations of deposited cues. 
\end{abstract}

\maketitle

\section{Introduction}

Searching efficiently for a target is a universal task encountered in a broad range of   natural and artificial processes, across multiple length and time scales \cite{Grebenkov2024}. Examples include, in increasing order of scales, chemical reaction kinetics \cite{Smoluchowski1918, Collins1949, Rice1985}, genetic transcription \cite{Berg1981, Gorman2008}, immune response mechanisms \cite{DeOliveira2016, Krummel2016, Levin2016, Fricke2016}, animal and human foraging \cite{Hassell1978, Traniello1989ForagingAnts, Hills2013}, search-and-rescue missions \cite{Drew2021, Lyu2023, Yedilkhan2025}, this list being far from exhaustive. Despite their very different natures, all these examples share the need for the searching agents to reach a target as fast as possible by tuning adjustable parameters, use external resources and develop implementable strategies to increase the rate at which they reach unexplored regions of space and thus increase the probability to locate their target quickly. 

Because of this broad relevance, understanding how to optimize search has become a central topic in statistical physics, and many mechanisms to improve search efficiency have been characterized over the past decades. Let us mention e.g. Lévy walks \cite{Edwards2007}, intermittent strategies \cite{Benichou2011}, stochastic resetting \cite{Chechkin2018, Wald2021, Olsen2026}, persistent motion \cite{Tejedor2012}, collective searches \cite{MejiaMonasterio2011, Meyer2024, Grebenkov2025}, trail-leaving walks \cite{Wysocki2026, Pineau2026}, non-Markovian dynamics \cite{Meyer2021, Barbier-Chebbah2024, Coghi2025, Dolgushev2025, Bremont2025, Netz2026b}, active particles \cite{Rupprecht2016, Basu2024, Jeon2024}, to name but a few. Many of these approaches require to modify the original dynamics of the searcher, which almost always implies a supply of energy or resources that allow for such modification of the dynamics. Although statistical measures of the overall search time remain the central quantities to optimize in the vast majority of these works, recent studies have sought to identify cost-effective strategies that reduce both the time required to find a target and the cost of implementing the search strategy. For example, resetting a searcher requires mechanical work \cite{Sunil2023, Tal-Friedman2025, Biswas2025}, hiring multiple searchers may cost energetic and material resources \cite{Meyer2025}, or even controlling the protocol itself may be costly \cite{Davis2026}. Most of these works assume however that resource availability is infinite: using it may be costly but it can always be done and one tries to minimize a relevant cost function. This is particularly true for non-Markovian search processes where the memory of the process is mediated by cues deposited along the path of the searcher, or by an alteration of the environment that can be sensed and interpreted at later times. This class of processes where agents interact with their own past trajectories has been at the center of recent studies and shown to be efficient search processes \cite{Altshuler2024, Singh2026, Wysocki2026,  Pineau2026}, but the ability of the searcher to produce environmental memory has always been considered unbounded.

In this work, we address a different but closely related question. Rather than assigning a continuous cost to resource consumption, we assume that the searcher possesses a finite amount of resource that can be used during the search. The objective is no longer to minimize a weighted combination of search time and cost, but rather to determine how a limited budget should be allocated in order to minimize the typical search time. This formulation is motivated by many realistic situations in which the available resource is fixed beforehand and cannot be replenished during the search \cite{Wolf1975, Ydenberg1994, Giang2024}. Only a few papers made a step towards questioning search strategies with finite resource availability in recent years \cite{Sato2010, Grebenkov2022, Linn2026}. 

The main problem that a finite resource budget raises is the following: how does one make sure that the finite budget is used as much of its full capacity as possible while not spending all of it too fast and searching for most of the remaining time without benefiting from it? In addition, how should one adjust the resource allocation strategy depending on the available information on the current state of the searcher? We therefore distinguish between two classes of processes. First, those where the searcher -- or an external agent that controls the resource usage -- knows at all times the current position of the walker, which we refer to as {\it informed} strategies. It can therefore condition the usage of the resource on it, or on the advancement of the process, and design a {\it spatial} protocol: whenever the searcher reaches a certain position for the first time, then it can use a certain pre-determined amount of the resource. In these strategies, resources should be used judiciously to exploit the available budget while minimizing the amount of unused resource remaining when the searcher reaches the target. Second, when neither the searcher nor any external controller knows the searcher’s current position, resource usage must follow a predetermined protocol. For such {\it blind} strategies, the optimal protocol should aim to maximize the probability that the resource budget is entirely used by the time the searcher reaches the target, while ensuring that the way it has been allocated does not hinder the search once the budget is exhausted and the search continues without additional resources. \\

To investigate this optimization problem, we start with a one-dimensional diffusive searcher which we provide with three different types of limited resources to speed up its search performance. First, we start with a reference process in which we apply an external pulling force and constrain the total work that can be expended by it before reaching the target, or equivalently the total heat dissipated. This is the most straightforward type of resource, as it keeps the process Markovian and requires standard techniques to be quantitatively described \cite{vanKampen1992StochasticProcesses, Redner2001}. However, applying an external force to the searcher requires an external controller, which is not always possible. Consequently, the second model we consider equips the searcher with a supply of directional flags that it can deposit along its path and that act as checkpoints redirecting it towards the target. This makes the process non-Markovian and effectively self-interacting, since the deposited flags alter the subsequent dynamics based on the searcher's previous positions. Still, this model assumes that the resource carries some information about the location of the target. We therefore consider, as a third model, isotropic chemotactic cues deposited by the searcher along its path which act as an effective repelling force. The search performance of such auto-chemotactic walkers has been recently investigated and found to be efficient \cite{Barbier-Chebbah2022, Wysocki2026, Pineau2026}, but always assuming an infinite supply of cue. In contrast, we consider here a finite supply and investigate how the resulting optimal search strategies are altered.

For all these processes, we consider a fixed target whose location is known, which reduces to an exit problem. However, there are multiple reasons for this choice. First, imposing the target location and optimizing the search strategy accordingly allows to uncover the main mechanisms that govern the optimal resource usage protocols. Second, analytical results are more accessible as introducing a probability distribution on the location of the target introduces nonlinearities in the expressions for first-passage times. Third, we discuss in the last section how our results are modified when we consider a target homogeneously distributed across a spatial domain, and find that the phenomenology for the fixed-target case survives. \\

\section{Problem Setting}

Let us start by introducing the general framework we consider and recall the free diffusion case as a reference case. A diffusive walker is initialized at $x=0$ on the semi-infinite line, i.e. we have a reflecting boundary at $x=0$ and open space for $x>0$, and we aim to study the time it requires to first reach the position $x=L$. This problem has long been solved and can be summarized as follows. The target survival probability $s(t)$, i.e. the probability that the absorbing boundary has not yet been reached at time $t$ is fully determined as 
\begin{equation}
    s(t) = \int_0^L dx c(x,t)
\end{equation}
where $c(x,t)$ is the associated concentration profile, which is the solution of the Fokker-Planck equation 
\begin{equation}
    \partial_t c(x,t) = D \partial_{x}^2 c(x,t)
\end{equation}
with boundary conditions $c(x,0) = \delta(x)$ and $c(L,t) = 0$, and diffusion constant $D$. This profile is known analytically and reads
\begin{equation}
    \label{eq:diff_concentration}
    c(x,t) = \frac{2}{L}\sum_{k=0}^\infty \cos\left(\frac{\lambda_k  x}{L}\right)e^{-\frac{\lambda_k^2}{L^2} Dt}
\end{equation}
where we have introduced $\lambda_k = (2k+1)\pi/2$ \cite{Redner2001}. The first-passage time distribution $\rho(t)$ is then found as $\rho(t) = -\dot{s}(t)$, the mean of which reads 
\begin{equation}
    \bar{\mathcal{T}}_{0} = \frac{L^2}{2D}   
\end{equation}
This diffusive mean first-passage time (MFPT) will be our reference point for the rest of the study. \\

Our goal is to improve this typical exit time given external {\it resources} that may accelerate the process are available to the searcher in a finite amount. In the following, we consider three models of such resources, and we investigate two types of strategies depending on whether the searcher or an external operator knows where the searcher is and can monitor the process at any point in time. If it can, the resource usage strategy is referred to as an {\it informed} strategy as the searcher can decide to use the resource based on its current state. If it cannot, the strategy is referred to as {\it blind} and the searcher uses its resources only following a predetermined time protocol, regardless of its current state.

\section{External pulling}

\subsection{Model}

We here consider as the available resource an external force $F$ that biases the motion towards the target and that can be applied to the searcher along its trajectory. The energetic cost associated to this force is simply the work performed by it until the absorbing boundary is reached. We therefore constrain it to be such that the total work budget is fixed, i.e. it cannot exceed a prescribed value $W$ at any point along the trajectory. Equivalently, we impose that the total heat dissipated until the end of the process is fixed and can never exceed its prescribed value $Q=W$ over the course of it. In other words, we do not allow for any credit for the resource budget at any point in the process.

\begin{figure*}
    \centering
    \includegraphics[width=\linewidth]{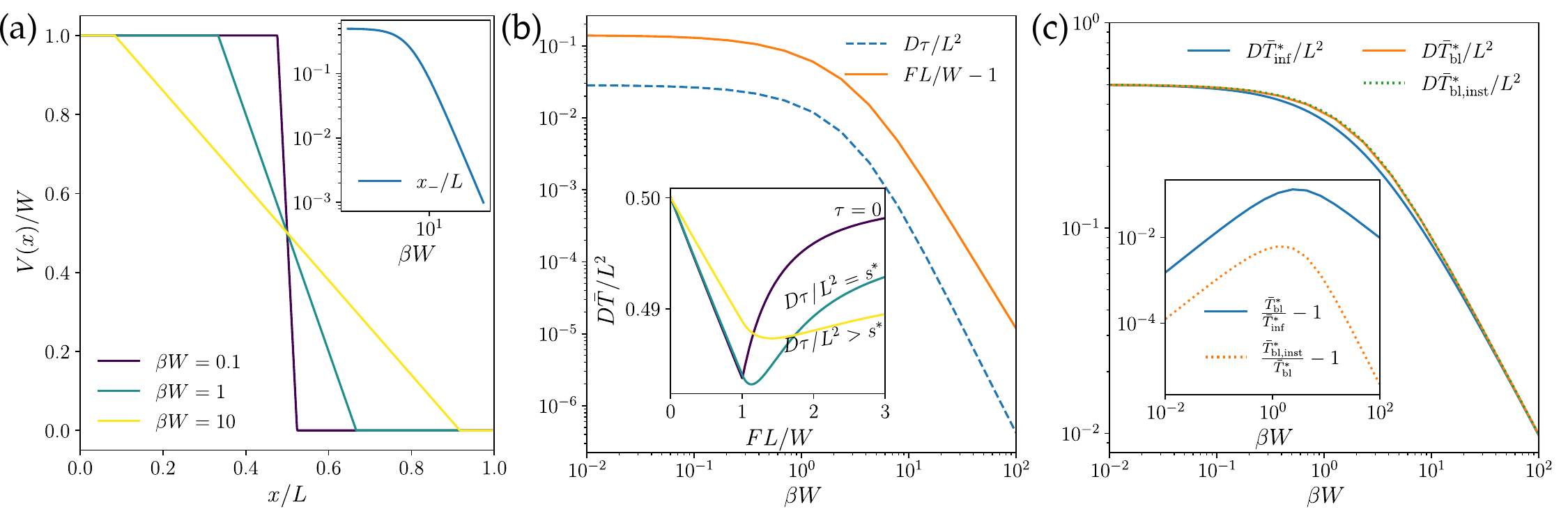}
    \caption{External pulling strategies. (a) Optimal force profiles for the informed strategy, for various values of the work budget $W$. The piecewise linear shape is symmetric around the center of the domain and is demarcated by the threshold positions $x_-$ and $x_+$, with $x_+ = L-x_-$, which is shown in the inset as a function of the available work budget $W$. (b) Optimal parameters for the blind strategies, i.e. the switching time $\tau$ and pulling force $F$, as a function of $W$. As $W$ increases, the switching time tends to zero and the optimal force becomes asymptotically close to $W/L$. The inset shows the MFPT as a function of the pulling force for various values of the switching time, namely instantaneous switch on, the optimal switching time, and a greater value. (c) Optimal values of the MFPT in the informed and blind strategies. The dotted line shows the near-optimal parameters $F=W/L$ and $\tau=0$ in the informed case. The inset shows the relative performance between the optimal informed and strategies (solid line) and between the near-optimal strategy and actually optimal one for the blind case (dotted line), which are both maximized for $\beta W \simeq 1$.}
    \label{fig:optimal_search_pulling}
\end{figure*}

\subsection{Informed strategy}
\label{sec:informed_pulling}

In the informed strategy, the searcher -- or an external observer -- can track the trajectory at any point in time and can therefore tune the force according to the current position of the searcher. We therefore consider a space-dependent, conservative force $F(x)  = -V'(x)$, where $V(x)$ is the associated potential profile. As the total work performed between $x=0$ and $x=L$ is given by the potential drop $V(0) - V(L)$, the work budget constraint is equivalent to setting $V(L) = 0$ and requiring $V(0) = W$. In addition, the requirement of no work credit makes the potential bounded between $0$ and $W$ over the full interval $[0,L]$. Given a potential profile $V(x)$, the MFPT to reach position $x=L$ is given by Kramers formula, reading \cite{Redner2001}
\begin{equation}
\label{eq:Kramers}
\bar{T}_\mathrm{inf} = \frac{1}{D} \int_{x_0}^L dx \int_0^x dy \,
e^{\beta [V(x) - V(y)]},
\end{equation}
where $\beta$ is the inverse temperature which we use to introduce the normalized work budget $w =\beta W$, and $x_0$ is the initial position which we set to $0$ in our case. To find the optimal profile that satisfies our constraint and minimizes this quantity, we first construct a piecewise potential $V^{(N)}(x; x_1, \cdots, x_{N})$ across $N+1$ regions demarcated by positions $x_k$, namely
\begin{equation}
    V^{(N)}(x; x_1, \cdots, x_{N}) = \frac{W}{N} \sum_{k=1}^{N} \Theta\left(x_k - x \right)
\end{equation}
where $\Theta$ is the Heaviside function. At each of the positions $x_k$, a potential drop of hight $W/N$ occurs such that for $x_k \leq x <x_{k+1}$ we have $V(x) = W\left(1-k/N\right)$.  This construction ensures the constraint on the work budget to be satisfied. The resulting MFPT reads
\begin{align}
    \label{eq:MFPT_discrete_rk}
    \bar{T}_\mathrm{inf} =& \frac{1}{2D}\sum_{k=0}^{N}\sum_{l=0}^N d_k d_l \rho^{|k-l|} 
\end{align}
where $\rho = e^{-w/N}$ and $d_i = x_{i+1}-x_i$ is the interval between potential jumps, with the convention $x_0 = 0$ and $x_{N+1} = L$. Note that the fixed length of the system imposes $d_N = L - \sum_{k=0}^{N-1}d_k$ such that Eq.~(\ref{eq:MFPT_discrete_rk}) can be rewritten only in terms of $\delta_i$ for $0\leq i\leq N-1$, which we aim to adjust in order to minimize $\bar{T}$:
\begin{align}
    \label{eq:MFPT_disc_pull}
    \bar{T}_\mathrm{inf} =& \sum_{k=0}^{N-1}\sum_{l=0}^{N-1}\frac{d_k d_l}{2D} \left(\rho^{|k-l|}-2\rho^{N-l}+1\right)\\
    &+\frac{L}{D}\sum_{l=0}^{N-1} d_l \left(\rho^{N-l} -1\right) + \frac{L^2}{2D} 
\end{align}
As this expression is quadratic in $d_k$, its gradient with respect to $d_i$ is linear and the optimization problem can be formulated as a matrix inversion problem. More specifically, we have
\begin{equation}
    \nabla \bar{T} = \mathsf{M} \delta - \mathrm{u}
\end{equation}
where  $(\nabla \bar{T})_i=\partial_{d_i} \bar T$ is the gradient of $\bar{T}$ with respect to the intervals $d_i$, $\mathsf{M}$ is a matrix and $\mathrm{u}$ is a vector defined as
\begin{align}
    \mathsf{M}_{ij} =& 1  - \rho^{N-j} - \rho^{N-i} + \rho^{|i-j|} \\
    \mathrm{u}_i =& L (1-\rho^{N-i}) 
\end{align}
where the indices $i$ run from $0$ to $N-1$. The optimal set of intervals $d_i$ is thus found by solving $\nabla \bar{T} = 0$, whose solution reads $d_i = \left( \mathsf{M}^{-1}\mathrm{u}\right)_i$. This turns out to be computable exactly, reading
\begin{align}
    d_i =&\frac{1 - e^{-w/N}(1-\delta_{i,0} )  }{N+1 - (N-1)e^{-w/N}} L
\end{align}
where $\delta_{i,j}$ is the Kronecker delta. We then take the continuous limit $N\to \infty$ and find the optimal potential $V^*(x)$ to be piecewise linear, i.e.
\begin{equation}
V^{*}(x) = W\times \left\{ 
\begin{tabular}{l l}
    $1$ &    , $x\leq x_-$ \\
    $\frac{x_+-x} {x_+-x_-}$ &   , $x_-< x\leq x_+$\\
    $0$ & , $x>x_+$
\end{tabular}
\right.
\end{equation}
where the boundaries of the non-zero force region are found as
\begin{equation}
\frac{x_\pm}{L} = \frac{1}{2}\left(1  \pm \frac{w}{2+w} \right)    
\end{equation}
For such an optimal force profile, the mean first-passage time is found as
\begin{equation}
\bar {T}_{\mathrm{inf}} = \frac{\bar{\mathcal{T}}_{0}}{1+\frac{w }{2}}.
\end{equation}
These results are illustrated in figure \ref{fig:optimal_search_pulling}a. They already highlight the advantage of concentrating the resource usage within a finite region rather than distributing it uniformly across the system. Note that the optimal profile can also be obtained by a standard Euler-Lagrange optimization using Karush-Kuhn-Tucker conditions to ensure the potential to be bounded between $0$ and $W$. However, the result for a finite value of the number of intervals $N$ will become useful in a later section of the paper.  


\subsection{Blind strategy} We now consider a situation in which the driving protocol cannot depend on the instantaneous position of the searcher. The controller cannot observe the state of the searcher nor monitor the progress of the process in real time. The only available control parameter is time itself. One must therefore decide \emph{a priori} when to apply a force and how strong this force should be. To model this situation, we assume that a constant force $F$ is switched on at a prescribed time $\tau$. The force is then maintained until the total work performed reaches a fixed budget $W$, after which it is switched off. Since a constant force performs work linearly with distance, the maximal displacement that can be induced by the force is $\Delta = \frac{W}{F}$.\\

We first consider the case $L>\Delta$. The searcher initially diffuses freely until time $\tau$, which yields three possible outcomes: (i) the searcher may have reached the target at an earlier time than $\tau$; (ii) it may have reached a position $x$ such that $x+\Delta \geq  L$, in which case the force is then applied until the particle reaches the target; (iii) it may have reached a position $x$ such that $x+\Delta < L$, in which case the force is then applied until the particle reaches position $x+\Delta$, where the work budget is exhausted and the force is then switched off. The searcher then resumes free diffusion until it eventually reaches the target at $x=L$. This results in a MFPT reading
\begin{equation}
\begin{aligned}
\bar T_{\mathrm{bl}} &=\int_0^L dx \int_0^\tau dt \, c(x,t)\\
& +\int_{L-\Delta}^L dx \,c(x,\tau)\bar{T}_F(x\to L)\\
&+ \int_0^{L-\Delta} dx \, c(x,\tau)\left[\bar{T}_F(x\to x+\Delta)+\bar{T}_0(x+\Delta\to L)\right]
\end{aligned}
\end{equation}
Each term here accounts for the three possible outcomes. Here $\bar{T}_F(x\to y)$ denotes the MFPT to reach position $y$ starting from $x$, under a constant force $F$, while $\bar{T}_0(x\to y)$ corresponds to the purely diffusive limit with $F=0$. It is obtained via Eq.~(\ref{eq:Kramers}) using a linear potential and reads 
\begin{equation}
\bar{T}_F(x\to y) = \frac{1}{D}\left[\frac{y-x}{\beta F} - \frac{e^{-\beta F x}-e^{-\beta F y}}{\beta^2 F^2}\right]
\end{equation}

In the opposite case $\Delta \ge L$, the force is in principle weak enough to bring the particle directly to the target before the work budget is exhausted and we therefore obtain
\begin{equation}
\bar T_{\mathrm{bl}} = \int_0^\tau dt \int_0^L dx \, c(x,t) + \int_0^L dx \,c(x,\tau)T_F(x \to  L).
\end{equation}

Introducing the normalized quantities $f = \beta FL$, $w = \beta W$ and $s = \tau D/L^2$, we obtain analytical expressions for these two cases, reading
\begin{widetext}
    \begin{align} \bar{T}_{\mathrm{bl}}(f<w) = &
\bar{\mathcal{T}}_0 - \frac{2L^2}{D}\sum_{k=0}^{\infty}
\frac{e^{-\lambda_k^2s}}{\lambda_k} \left[(-1)^k  - \frac{1}{f^2 + \lambda_k^2}\left(\frac{f}{\lambda_k}  + e^{- f}(-1)^k\right) 
\right]\\
\bar{T}_{\mathrm{bl}}(f\geq w) =& 
\bar{\mathcal{T}}_0 - \frac{2L^2}{D}\sum_{k=0}^{\infty}
\frac{e^{-\lambda_k^2s}}{f^2 +\lambda_k^2} 
\Bigg\{ \frac{1 - w- e^{-w}}{f} +  \left[ (-1)^k \left(1 + \frac{f^2}{\lambda_k^2}\right) \sin\left(\frac{\lambda_k w}{f} \right) -  \frac{f w}{\lambda_k^2} \right]  \nonumber \\
&\qquad \qquad \qquad \qquad \qquad \qquad + \frac{(-1)^k}{f} \left[e^{-f}-1 - \frac{f^2}{\lambda_k^2}\right] \left( \sin\left(\frac{\lambda_k w}{f} \right) + \frac{f}{\lambda_k} \left(\cos\left(\frac{\lambda_k w}{f} \right)-1\right)  \right) 
\Bigg\}
 \end{align}
\end{widetext}
Optimizing over the protocol parameters, namely $\tau$ and $f$, we obtain that the optimal strategy requires to start applying a force $F = W/L + \delta f$ with $\delta f \ll W/L$ at a very short time $\tau$, i.e. $\tau D/L^2 \ll 1$, yet not zero, as shown in figure \ref{fig:optimal_search_pulling}b. However, the resulting optimal MFPT $\bar{T}^*_{\mathrm{bl}}$ differs from the one reached for $F = W/L$ and $s = 0$ by a negligible amount, which reads
\begin{equation}
    \bar{T}_{\mathrm{bl}}(f=w, s=0) = 2\bar{\mathcal{T}}_0\frac{e^{-w} -1+w }{w^2} 
\end{equation}
At most, the fine-tuned $\bar{T}^*$ differs from $\bar{T}(f=1, s=0)$ by less than a percent, for $\beta W = 1$, as shown in figure \ref{fig:optimal_search_pulling}b. In other words, when the controller has no information about the position of the searcher, it is only negligibly advantageous to wait before applying the force. In the nearly optimal case of an instantaneous force switch, the available work budget is exhausted exactly when the particle reaches the absorbing boundary. This result differs markedly from the informed pulling protocol discussed previously where the optimal strategy concentrates the driving force within a narrow spatial region. In contrast, optimal blind pulling favors a uniform and continuous expenditure of the available work across the whole interval. While the optimal informed strategy performs better than the blind one, its resulting MFPT is lower by only at most $14\%$, this maximum value being reached for $\beta W = 2.69$, see figure \ref{fig:optimal_search_pulling}c. It demonstrates that a blind usage of the work budget can still perform significantly well in comparison to an informed usage of it, if properly optimized, and intermediate values of the work budget are those for which the search performances differ the most. On one end, a low budget can only negligibly improve the search time such that optimizing the resource usage does not improve it significantly regardless of the strategy. On the other end, a high budget is such that the details of the resource usage does not matter much as the resource availability is high. \\

\section{Flag deposition}

Pulling a searcher via an applied force requires external control. This may not be possible in many realistic processes. Instead, the searcher may have access to cues that it can deposit along its path to effectively induce memory of its trajectory, as observed in e.g. insect foraging \cite{Traniello1989ForagingAnts} or robotic exploration \cite{Liu2021}. We here introduce a model of {\it directional} flags that a searcher can deposit along its path. Similarly to the application of an external force, the flags are designed to push the walker towards the target. They therefore contain directional information and are not left-right symmetric.\\

\begin{figure*}
    \centering
    \includegraphics[width=\linewidth]{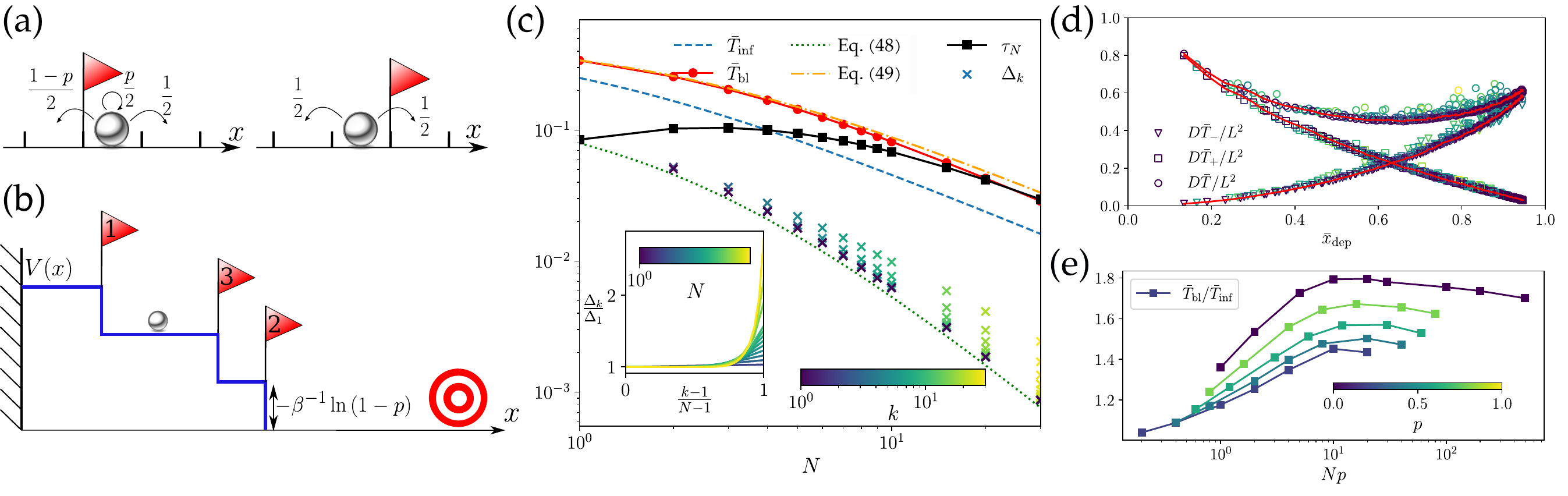}
    \caption{Flag deposition strategies. (a) Sketch of the lattice model. The flag acts as partially reflecting objects which redirects the walker towards the target, i.e. greater $x$ values. (b) The lattice model can be mapped in the continuous limit to a diffusive walk in a potential made of discontinuous, downward jumps at the locations of flags. In the blind deposition scheme, for $p<1$, the spatial ordering of flags does not necessarily follow the chronological order of depositions. (c) Optimal blind strategies in the totally reflecting case $p=1$. The deposition intervals $\Delta_k$ are almost constant as shown by the crosses which almost all lie on top of each other for a given $N$, and in the inset showing the almost constant unity ratio $\Delta_1/\Delta_k$ as a function of $(k-1)/(N-1)$ for increasing values of $N$. The optimal deposition interval $\Delta^*$ and corresponding MFPT $\bar{T}^*_{bl}$ is well approximated by Eq.~(\ref{eq:approx_Tbl_flags}) (dotted lines) obtained under the assumption $\Delta_k = \Delta$, with deviations growing with larger values of $N$. (d) Blind MFPT $\bar{T}_{\mathrm{bl}}$ and pre- and post-depletion times $\bar{T}_-$ and $\bar{T}_+$ as a function of the mean depletion position $\bar{x}_{\mathrm{dep}}$. Each point represents a random set of deposition times $\tau_k$ while the solid line shows the result for a constant deposition interval $\tau_k = k\Delta$, which goes through the overall minimum of $\bar{T}_{\mathrm{bl}}$ up to numerical accuracy. (e) Relative performance $\bar{T}_{bl}/\bar{T}_{inf}$ between optimal blind and informed strategies as a function of $Np$ for various values of $p$. The ratio exhibits a maximum value as a function of $N$ but decreases towards a plateau.}
    \label{fig:flags}
\end{figure*}

\subsection{Model}

Let us first introduce the space- and time-discrete version of the model before considering the continuous limit. Consider a one-dimensional lattice of size $N_L$, with lattice constant $a$. The walker evolves on lattice sites and its position at time step $n$ is noted $i_n$. The flags, however, are placed on lattice bonds, and the number of flags on the bond between sites $i-1$ and $i$ is noted $f_i$. At time step $n$, the probabilities for the walker to jump from its current site $i_n$ to either sites $i_n\pm 1$, or to remain at site $i_n$ are defined as 
\begin{gather}
    p_{i_n\to i_n-1} = \frac{(1-p)^{f_{i_n}}}{2}, \qquad p_{i_n\to i_n} = \frac{1-(1-p)^{f_{i_n}}}{2}, \nonumber \\ 
    p_{i\to i+1} = \frac{1}{2}
\end{gather}
Here the probability $p$ acts as an interaction strength: for $p=1$, the flags are purely reflecting and act as effective walls; for $p=0$, the flags are not sensed at all by the searcher which simply performs standard diffusion. A sketch of the model is depicted in figure \ref{fig:flags}a.

To obtain the continuous limit of this model, we note that between two consecutive flags, the process is a purely diffusive random walk. Probability conservation and continuity of the probability current implies that the flags essentially act as potential jumps such that the overall process can be mapped on a diffusive walk in a potential profile  $V(x; x_1,\cdots, x_N)$, where $x_k$ is the position of the $k$-th flag, reading 
\begin{equation}
    V(x; x_1,\cdots, x_N) =  - \beta^{-1} \ln (1-p) \sum_{k=1}^N \Theta(x_k-x)
\end{equation}
where $\Theta(x)$ is the Heaviside function, as illustrated in figure \ref{fig:flags}b. The process we consider, however, is such that the flags are not placed at fixed locations from the start of the process. In fact, the searcher has a finite supply of $N$ flags that can be deposited throughout the trajectory. The effective potential is therefore modified whenever a new flag is deposited. As in the previous section, two distinct optimization strategies can be considered. In the informed case, the searcher knows its current position at all times and can thus decide to deposit the flags on the basis of this information. In the blind case, the searcher cannot track its position and therefore cannot condition the flag deposition on it. The deposition strategy therefore consists of a temporal protocol that is prescribed from the start of the process.

\subsection{Informed deposition}

We first introduce the informed deposition protocol in the discrete setting. Before the process begins, we first identify a set of sites ${j_1, \cdots, j_N}$ where flags are going to be deposited. When the searcher reaches one of the sites $j_k$ for the first time, it deposits a flag on the bond between sites $j_k$ and $j_{k}-1$. If the site $j_k$ has been visited at an earlier time, no new flag is deposited. Because the effect of a flag is localized at a lattice bond and only affects transitions from $j_k$ to $j_{k}-1$ -- and not from $j_{k-1}$  to $j_k$ -- this deposition scheme is such that the walker explores a predetermined set of flag locations, or equivalently a predetermined potential profile in the continuous limit. Consequently, optimizing this deposition strategy is equivalent to optimizing a Kramers-like MFPT -- that is, of the form of Eq.~(\ref{eq:Kramers}) -- with respect to an effective potential made of discrete downward steps. This is precisely what we performed in section \ref{sec:informed_pulling} as the discretized potential $V^{(N)}$ was optimized before taking the limit $N\to \infty$. The only difference here is the size of the potential drops, which are now $- \beta^{-1} \ln (1-p)$ instead of $W/N$. We can therefore use Eq.~(\ref{eq:MFPT_disc_pull}) for the MFPT and substitute $1-p$ in place of the exponential factors $e^{-w /N}$. The minimization can then be carried out analytically, yielding the optimal interval lengths $d_i$ between flags $i$ and $i+1$
\begin{equation}
d_i=\frac{p + (1-p)\delta_{i0}}{2+(N-1)p} L
\end{equation}
where $\delta_{ij}$ is again the Kronecker delta. This solution corresponds to a configuration in which the first flag is deposited at a larger distance from the origin, while the remaining flags are spaced uniformly over shorter intervals. For this optimal arrangement, the MFPT becomes
\begin{equation}
\bar{T}_{\mathrm{inf}} = \frac{2-p}{2+(N-1)p}\bar{\mathcal{T}}_0
\end{equation}
Importantly, the MFPT decreases asymptotically as $1/N$, showing that the search efficiency can be significantly enhanced by increasing the number of deposited flags. Similarly to the external pulling process, optimal informed strategies require the deposition of flags to be concentrated in a region of space whose width becomes smaller as the resource supply decreases. 

\subsection{Blind deposition}

In the blind deposition problem, the searcher cannot control where the flags are deposited and may only determine {\it when} to release them according to a predetermined protocol. This gives rise to two key features that distinguish the blind setting substantially from the informed case. First, because the searcher has no information about its current position, it may reach the target before the supply of flags is depleted. Conversely, the supply may be depleted before the target is reached, in which case both the searcher’s position at the time of depletion and the locations of previously deposited flags influence its subsequent dynamics. This contrasts with the external pulling process, in which blind strategies switch off the force once the work budget is exhausted. In the present setting, the flags remain at the locations at which they were deposited. Second, whereas in the informed case the spatial ordering of the flags necessarily follows their chronological order of deposition, in the blind setting a flag deposited at a later time may be positioned between two flags deposited earlier. Consequently, the searcher does not evolve in a fixed potential landscape, unlike th informed protocols. 

To characterize the pre- and post-depletion stages, we introduce the depletion time $\tau_{\mathrm{dep}}$, i.e. the time at which the last flag in the supply is deposited. The target survival probability can naturally be decomposed into two components, namely
\begin{equation}
    S(t) = S_-(t)\Theta(\tau_{\mathrm{dep}}-t) + S_+(t)\Theta(\tau_{\mathrm{dep}}-t)
\end{equation}
where $\Theta$ is the Heaviside function. The MFPT can therefore be obtained as 
\begin{align}
    \bar{T}_{\mathrm{bl}} =& \int_0^{\tau_{\mathrm{dep}}} dt S_-(t) + \int_0^{\tau_{\mathrm{dep}}} dt S_+(t)  \\
    \label{eq:Tm_Tp}
    =& \bar{T}_-+\bar{T}_+
\end{align}
Here $\bar{T}_-$ and $\bar{T}_+$ are the expected values of the random variables $T_-$ and $T_+$, respectively, which for each trajectory are defined in terms of the search time $T$ as $T_- = \min\left\{T, \tau_{\mathrm{dep}} \right\}$ and $T_+ = \max\left\{0, T - \tau_{\mathrm{dep}} \right\}$. An optimal strategy should be one that balances these two competing contributions.

\subsubsection{Total reflection} 
We start our analysis by considering the case $p=1$ where the searcher is always reflected against the flag, whose mathematical structure allows to derive useful analytical results. Let $\tau_k$ be the time at which the $k$-{th} flag is deposited, and $\Delta_k = \tau_{k}-\tau_{k-1}$ be the time interval between two consecutive depositions. The total MFPT $\bar{T}_{\mathrm{bl}}(L)$ -- whose dependence on the system size is made explicit for reasons that will be apparent later -- is expressed as the time integral of the target  survival probability $s(L, \Delta_1, \cdots, \Delta_{N}, t)$, that we decompose into a piecewise function:
\begin{align}
    \bar{T}_{\mathrm{bl}}(L) =& \int_0^\infty dt s(L, \Delta_1, \cdots, \Delta_{N}, t) \nonumber \\
    =& \sum_{n=0}^N \int_{\tau_n}^{\tau_{n+1}} dt s_n(L, \Delta_1, \cdots, \Delta_{n}, t) \nonumber\\
    =& \sum_{n=0}^N \int_{0}^{\Delta_{n+1}} dt S_n(L, \Delta_1, \cdots, \Delta_{n}, t)
\end{align}
Here $S_n(L, \Delta_1, \cdots, \Delta_{n}, t) = s_n(L, \Delta_1, \cdots, \Delta_{n}, t-\tau_n)$ is the probability that the target has not yet been reached at time $t$, for a process in a domain of size $L$ with flags deposited at time intervals $\Delta_1, \cdots, \Delta_{n}$. Because the case $p=1$ is such that every flag deposition acts as an effective shrinkage of the search domain, the partial survival probabilities $S_n$ obey the recursive relation
\begin{align}
    \label{eq:recursive_Sn}
    S_{n+1}&(L,\Delta_1,\cdots, \Delta_{n+1},t) \nonumber \\
    &= \int_0^L dx c(L,x,\Delta_1) S_{n}(L-x,\Delta_2,\cdots, \Delta_{n},t)
\end{align}
where we have made explicit the dependence on $L$ of the concentration profile introduced in Eq.~(\ref{eq:diff_concentration}). After a time $\Delta_1$, the trajectories that have not yet reached the target have reached a position $x$ distributed according to $c(L,x,\Delta_1)$. From there, a flag is deposited, and the process resumes as the original process in a smaller region of size $L-x$, whose partial survival probability is now $S_{n}(L-x,\Delta_2,\cdots, \Delta_{n},t)$. The initial term for this recursive relation is simply the flag-less survival probability, i.e. $S_{0}(L,t) = \int_0^L dx c(L,x,t)$. Now, we note that the concentration profile $c(L,x,t)$ is of the form $  c(L,x,t) = L^{-1}\tilde{c}\left(x/L, Dt/L^2 \right)$, which can be used to simplify the recursion relation. To do this, we first make the ansatz $S_n(L, \Delta_1, \cdots, \Delta_{n}, t) = \tilde{S}_n\left(\frac{D\Delta_1}{L^2},\cdots, \frac{D\Delta_n}{L^2},\frac{Dt}{L^2}\right)$ which we know holds for $n=0$, and we introduce the following notations
\begin{align}
    \Psi^{(n)}_{v_1,\cdots, v_{n}, v}(u) =&   \tilde{S}_n\left(\frac{v_1}{u^2},\cdots, \frac{v_n}{u^2},\frac{v}{u^2}\right) \\
    \gamma_{v}(u) =&  \tilde{c}(u,v)
\end{align}
The recursion relation (\ref{eq:recursive_Sn}) can therefore be written as
\begin{align}
    \label{eq:recursive_Psin}
     \Psi^{(n+1)}_{v_1,\cdots, v_{n+1}, v}(u) =&  \int_0^u du' \gamma_{v_1}\left(u-u' \right) \Psi^{(n)}_{v_1,\cdots, v_{n}, v}(u')
\end{align}
evaluated at $u=1$, and with $v_k = D\Delta_k /L^2$. The associated, normalized MFPT $\mathfrak{T}_{v_1, \cdots, v_N}(u)$ is defined as 
\begin{align}
    \mathfrak{T}_{v_1, \cdots, v_N}(u) =&  \sum_{l=0}^{N} \int_0^{v_{l+1}} dv \Psi^{(l)}_{v_1, \cdots, v_l, v}(u)
\end{align}
Because of the convolution product in Eq.~(\ref{eq:recursive_Psin}), the rescursive relation becomes a simple product in Laplace space. Using the functional form of the concentration profile $c(x,t)$, i.e. Eq.~(\ref{eq:diff_concentration}), we finally obtain after some algebra
\begin{align}
    \label{eq:laplace_T_z}
    \hat{\mathfrak{T}}_{v_1, \cdots, v_N}(z) = \sum_{l=0}^{N}  &\left[\left( \frac{1}{z^3} -2 \sum_{j=0}^\infty (-1)^j J(2\lambda_j z, v_{l+1}) \right) \right. \nonumber  \\
    &\qquad \times \left. \prod_{k=1}^l \left( \sum_{i=0}^\infty \frac{2z e^{-\lambda_i^2 v_k}}{z^2 + \lambda_i^2}\right)  \right] 
\end{align}
where 
\begin{align}
     J(z,v) = \int_0^\infty du   e^{-z u-\frac{v}{u^2}}     u^2
\end{align}
and $\hat{f}(z) = \int_0^\infty dt e^{-zt}f(t)$ is the Laplace transform operation. Transforming back to real time space, the MFPT is then found as
\begin{equation}
    \bar{T}_{\mathrm{bl}}(L) = \frac{L^2}{D}\mathfrak{T}_{\frac{D \Delta_1}{L^2},\cdots, \frac{D \Delta_N}{L^2}}(u=1)
\end{equation}
Although this expression cannot be further simplified in general, it can be used to implement efficient gradient descent algorithm. We can in fact evaluate Eq.~(\ref{eq:laplace_T_z}) for arbitrary values of $z$ and inverse transform it to evaluate $\mathcal{T}_{v_1, \cdots, v_N}(u=1)$, instead of computing nested integrals involved in the original recursion relation (\ref{eq:recursive_Sn}) for $S_n$. This allows to probe relatively large values of $N$ and find the optimal set of deposition intervals $\Delta_k$ that minimize $\bar{T}$. 

As shown in Fig.~\ref{fig:flags}c, the optimal strategy consists of deposition intervals that are nearly equal for a given $N$. More precisely, the ratio $\Delta_k/\Delta_1$ remains close to unity for most values of $k$, with noticeable deviations occurring only for $k$ close to $N$. In particular, $\Delta_N/\Delta_1$ increases with $N$, but this effect becomes appreciable only for sufficiently large $N$. However, for processes involving many flags, the contribution of the final flags to the overall search time becomes increasingly negligible, since a substantial fraction of trajectories reach the target before the flag supply is depleted. Consequently, fine-tuning the deposition intervals for large $k$ when $N$ is also large has only a negligible effect on the overall search time.\\

If we impose such a constant deposition interval, i.e. $\Delta_k = \Delta$, the mean first-passage time can be worked out in the limit $v\ll 1$ and $v\gg 1$. Let us first rewrite 
\begin{align}
    \label{eq:Tv_constDelta}
    \mathfrak{T}_{v}(u) = \sum_{l=0}^{N-1} &\mathfrak{T}^{(l)}_v(u) + \mathfrak{T}^{(N)}_\infty (u)
\end{align}
with
\begin{align}
    \mathfrak{T}^{(l)}_v(u) =& \left[g_{v} *  \underbrace{\gamma_v * \cdots *  \gamma_v}_{l \text{ times}}\right](u), \\
    g_{v}(u)=&  \int_0^{v} dw \Psi^{(0)}_w\left(u \right) \nonumber\\
    =&  \frac{u^2}{2} \left[ 1 - 4\sum_{j=0}^\infty \frac{(-1)^j}{\lambda_j^3}e^{-\frac{\lambda_j^2}{u^2}v}\right]
\end{align}
Note that the long list of indices in the notation of all quantities involved here has been replaced by a single index since all introduction intervals are now the same. 
In the limit $v\gg 1$, the terms $l=0,1$ in Eq.~(\ref{eq:Tv_constDelta}) are dominant, within which the term $k=0$ in Eq.~(\ref{eq:diff_concentration}) used in $\gamma_v$ is also dominant. Keeping only these contributions leads to 
\begin{align}
    \bar{T}_{\mathrm{bl}} =&  \bar{\mathcal{T}_0} \left[1 - \frac{4(4 - \pi)}{\pi^3} e^{-\frac{\pi^2}{4}v} \right] + O\left( e^{-\frac{9\pi^2}{4}v} \right) 
\end{align}
The MFPT therefore approaches the resource-free limit $\bar{\mathcal{T}_0}$ exponentially fast from below. This limit does not depend on $N$ as it practically corresponds to a scenario where only the first flag effectively contributes to the search time. 

The limit $v\ll 1$ requires more care. We first note that we can rewrite $\gamma_v(u)$ using Poisson summation as a sum of Gaussians centered at $u=0,\pm 2, \pm4, \cdots$, whose widths are proportional to $\sqrt{v}$, i.e.
\begin{equation}
    \label{eq:gamma_gauss}
    \gamma_v(u) = \frac{1}{\sqrt{\pi v}}\sum_{m=-\infty}^\infty (-1)^m e^{-\frac{(u-2m)^2}{4v}}
\end{equation}
The leading term for $0\leq u \leq 1$ is the one for $m=0$, the other terms being exponentially negligible as $v \to 0$. In addition, we have $g_v(u) = v+ O(v^2)$ which therefore implies
\begin{equation}
    \mathfrak{T}^{(l)}_v(u) = v + O(v^{3/2})
\end{equation}
The first sum in $\mathcal{T}_{v}(u)$ thus results into $Nv$: this simply corresponds to the total time spent to deposit all flags, assuming all trajectories reach the target after depletion of the flag supply, which is asymptotically true as $v\to 0$. For the second term, we first remark that $g_\infty(u) = u^2/2$ and $\mathfrak{T}^{(N)}_\infty(u)$ is of the form  
\begin{equation}
    \mathfrak{T}^{(N)}_\infty(u) = \int_0^u dw \frac{(1-w)^2}{2} \Gamma_v(w)
\end{equation}
where $\Gamma_v(w)$ is the convolution product of $N$ times $\gamma_v$, i.e. it is the probability distribution of the sum of $N$ independent random variables, each distributed according to $\gamma_v$, which itself is a sum of Gaussians. For $u=1$, the term $m=0$ is exponentially dominant as $v\ll 1$, such that $\Gamma_v(w)$ is the probability distribution of the sum of $N$ random variables, each distributed according to the half-Gaussian $\gamma_v(u) = e^{-\frac{u^2}{4v}}/\sqrt{\pi v}$, whose mean is found as $2\sqrt{v/\pi}$ and second moment as $2v$. Because these distributions are narrow around $w=0$, we can approximate the integration domain $[0,1]$ by the full domain $[0,\infty)$ such that 
\begin{align}
     \mathfrak{T}^{(N)}_\infty(u=1) =& \frac{1}{2}\mathbb
    E_{\Gamma_v}\left[(1-w)^2\right]\nonumber\\
    =& \frac{1}{2} - 2N\sqrt{\frac{v}{\pi}} + Nv + 2N(N-1)\frac{v}{\pi}
\end{align}
where we have used $\mathbb{E}_{\Gamma_v}\left[w\right] = N\mathbb{E}_{\gamma_v}\left[w\right] $ and $\mathbb{E}_{\Gamma_v}\left[w^2\right] = N\mathbb{E}_{\gamma_v}\left[w^2\right] + N(N-1) \mathbb{E}_{\gamma_v}\left[w\right]^2$. All contributions together, we obtain
\begin{align}
    \bar{T}_{\mathrm{bl}} =& \mathcal{T}_0  - 2N\sqrt{\frac{L^2 \Delta}{D \pi }} + 2N\left(1 + \frac{N-1}{\pi}\right) \Delta \nonumber \\
    &+ O\left(\left(\frac{D\Delta}{L^2} \right)^{3/2}\right)
\end{align}
This low-$\Delta$ expansion already exhibits a minimum for an optimal deposition interval $\Delta^*$ reading
\begin{equation}
     \Delta^* = \frac{L^2}{D} \frac{\pi}{4(N-1+\pi)^2}
\end{equation}
This results in an optimal search time reading
\begin{equation}
    \label{eq:approx_Tbl_flags}
    \bar{T}_{\mathrm{bl}}^* = \bar{\mathcal{T}}_0 \frac{\pi - 1}{N-1+\pi}
\end{equation}
This approximation is remarkably close to the result obtained from numerical optimization, as shown in Fig.~\ref{fig:flags}, particularly for small and intermediate values of $N$. It predicts that the optimal deposition interval scales as $N^{-2}$, implying that the corresponding depletion time scales as $N^{-1}$, in the same way as the mean search time. 

In addition, the mean position of the last deposited flag can be made explicit. Since the flags are deposited at intervals $\Delta^*$, its mean position is given by $N$ times the average distance traveled by a freely diffusing particle on the semi-infinite line during a time $\Delta^*$, which is known to be $\sqrt{4D\Delta/\pi}$. Thus, for the optimal deposition interval $\Delta^*$, the mean position $\bar{x}_F$ of the last flag is
\begin{equation}
\bar{x}_F = N \sqrt{\frac{4D\Delta^*}{\pi}}
= \frac{N}{N-1+\pi}L,
\end{equation}
so that $\bar{x}_F\to L$ as $N\to\infty$. In this limit, the last flag is therefore deposited, on average, precisely when the target is reached. This is reminiscent of the optimal blind strategy for the external pulling process, in which the work budget is exhausted precisely at the time the target is reached. Unlike the informed protocol, Resource is consumed at a steady pace and tries to ensure the supply to be used at most of is capacity as the searcher reaches the target. \\

We can now compare the optimal MFPT for the blind case using the approximation (\ref{eq:approx_Tbl_flags}) to the optimal one in the informed case. In particular we find 
\begin{equation}
    \frac{\bar{T}^*_{\mathrm{bl}}}{\bar{T}^*_{\mathrm{inf}}} =  \frac{(N+1)(\pi - 1)}{N+\pi-1} 
\end{equation}
This ratio is not only always larger than 1 but also monotonically increases with $N$ towards the asymptotic value $\pi-1$. In other words, the blind strategy does not become asymptotically as good as the informed strategy in the limit of large flag supply, in contrast to the external pulling process. The robustness of this prediction is numerically tested for partially reflecting flags in the following.

\subsubsection{Partial reflection}

We now consider the case $p<1$, where the searcher can cross the flag with non-zero probability. Here, the convolutional structure of the equations is not preserved because a flag can in principle be deposited at a location closer to the origin than a flag deposited earlier. Eq.~(\ref{eq:laplace_T_z}) therefore does not hold. We can still write the survival probability in a recursive way, namely
\begin{widetext}
    \begin{equation}
    s_n(L, \tau_1, \cdots, \tau_{n}, t) = \int_0^L dx_1 \cdots \int_0^L dx_{N+1} 
    \chi_n(\left\{x_k, \tau_k \right\}_{k\leq N}) c_n(x_{n}, x,  t-\tau_{n}; \left\{x_k\right\}_{k\leq n}) 
\end{equation}
\end{widetext}
where the kernels $\chi_n$ is recursively defined as 
\begin{align}
    \chi_{n}&(\left\{x_k, \tau_k \right\}_{k\leq n}) \nonumber \\
    &= \chi_{n-1}(\left\{x_k, \tau_k \right\}_{k\leq n-1}) c_{n}(x_{n-1}, x_n, \Delta_{n}; \left\{x_k\right\}_{k\leq N}) 
\end{align}
and $c_j\left(x, x', t; \left\{x_k\right\}_{k\leq j}\right)$ is the concentration profile for a searcher initialized at position $x$ with $j$ flags at positions $x_1,\cdots, x_j$ to be at position $x'$ at time $t$, with absorbing boundary at $x=L$. The mean first passage time is then found as $ \bar{T}_{\mathrm{bl}} = \sum_{n=0}^N \int_{\tau_n}^{\tau_{n+1}} dt s_n(t) $, with the convention $\tau_0=1$ and $\tau_{N+1} = \infty$. A compact expression for $\bar{T}_\mathrm{bl}$ is only accessible for $N=1$, reading 
\begin{equation}
    \bar{T}_\mathrm{bl} =  \frac{L^2}{2D} \left[1+8p\sum_{k=0}^{\infty}  e^{-\frac{\lambda_k^2}{L^2}  D\tau_1}\left( \frac{1}{2\lambda_k^2}-\frac{(-1)^k}{\lambda_k^3} \right)\right]
\end{equation}
As expected, both limiting cases, $\tau=0$ and $\tau\to\infty$, recover the purely diffusive result $\bar T_\mathrm{bl}=\bar{\mathcal{T}}_0$, since the flag is ineffective in either limit: it is deposited either immediately, before the searcher has had time to explore the system, or sufficiently late that it has essentially no influence on the trajectory. For the single-flag process, the minimum of $\bar T$ is independent of $p$, since $p$ enters only as an overall prefactor. The optimal deposition time is found to be $\tau_1^*\simeq0.17,\bar{\mathcal{T}}_0$, which is substantially shorter than the mean exit time in the absence of flags. Consequently, only about $4\%$ of trajectories reach the target before the flag is deposited. Among the trajectories for which the flag is deposited before the target is reached, its mean position is $x_1=0.31L$, placing it considerably closer to the origin than the optimal position obtained in the informed strategy.

For $N>1$, we use numerical simulations to determine the optimal deposition strategy. For given $N$ and $p$, we select random deposition times $\left\{\tau_k \right\}_{1\leq k \leq N}$ and characterize their regularity through the constantness parameter $C = (N-1)^{-1} \sum_{k=1}^{N}\left| \frac{\Delta_k}{\bar{\Delta}}-1  \right|/2$, 
where $\bar{\Delta} = N^{-1}\sum_{k=1}^{N}\Delta_k$ is the mean interval between consecutive deposition times. Uniform deposition intervals give $C=0$, whereas $C=1$ when all $\Delta_k$ but one vanish, corresponding to deposition concentrated around a single instant. For each set of $N$, $p$, and $\left\{\tau_k \right\}_{1\leq k \leq N}$, we perform Brownian dynamics simulations and measure the pre- and post-depletion times $\bar{T}_-$ and $\bar{T}_+$ as defined in Eq.~(\ref{eq:Tm_Tp}), as well as the mean depletion position $\bar{x}_{\mathrm{dep}}$, which for each trajectory is defined as the position of the last deposited flag if $T\leq \tau_{\mathrm{dep}}$ and $x_{\mathrm{dep}} = L$ if $T> \tau_{\mathrm{dep}}$. We find that the MFPT for a constant deposition rate   $\tau_k=k\Delta$ reaches the overall minimum, within numerical accuracy, as shown in figure \ref{fig:flags}d, similarly to the case of totally reflecting flags. 

Plotting the mean pre- and post-depletion times as functions of the mean depletion position $\bar{x}_\mathrm{dep}$ further shows that the optimal strategy balances the two stages, with the two remaining comparable, as shown in figure \ref{fig:flags}d. Depletion should therefore occur sufficiently close to the target to limit the duration of the post-depletion stage, while avoiding an excessively long pre-depletion stage.

The optimal strategies retain the same asymptotic scaling as in the $p=1$ case: the mean first passage time decays as $N^{-1}$, while the deposition interval scales as $N^{-2}$ in the limit $N\to\infty$. Also, we show in figure \ref{fig:flags}e that the relative performance of the blind search compared to the optimal informed case initially grows with the number of flags $N$, then reaches a maximum and eventually settles to a plateau as $N\to\infty$, as it was already predicted for the totally reflecting flags.

\section{Auto-chemotaxis}

The model of flags presented in the previous section carries information about the location of the target since they always redirect the searcher towards it. The dynamics of the process is therefore not left-right symmetric. We now investigate searches where such helpful symmetry breaking is no longer present. In particular, we consider an auto-chemotactic process, where an agent produces itself a continuous cue along its path that exerts an effective repelling force on it, which is observed in biological systems, e.g. in immune cell mobility \cite{Krummel2016} or insect foraging \cite{Hassell1978Foraging}.

\begin{figure*}
    \includegraphics[width=\linewidth]{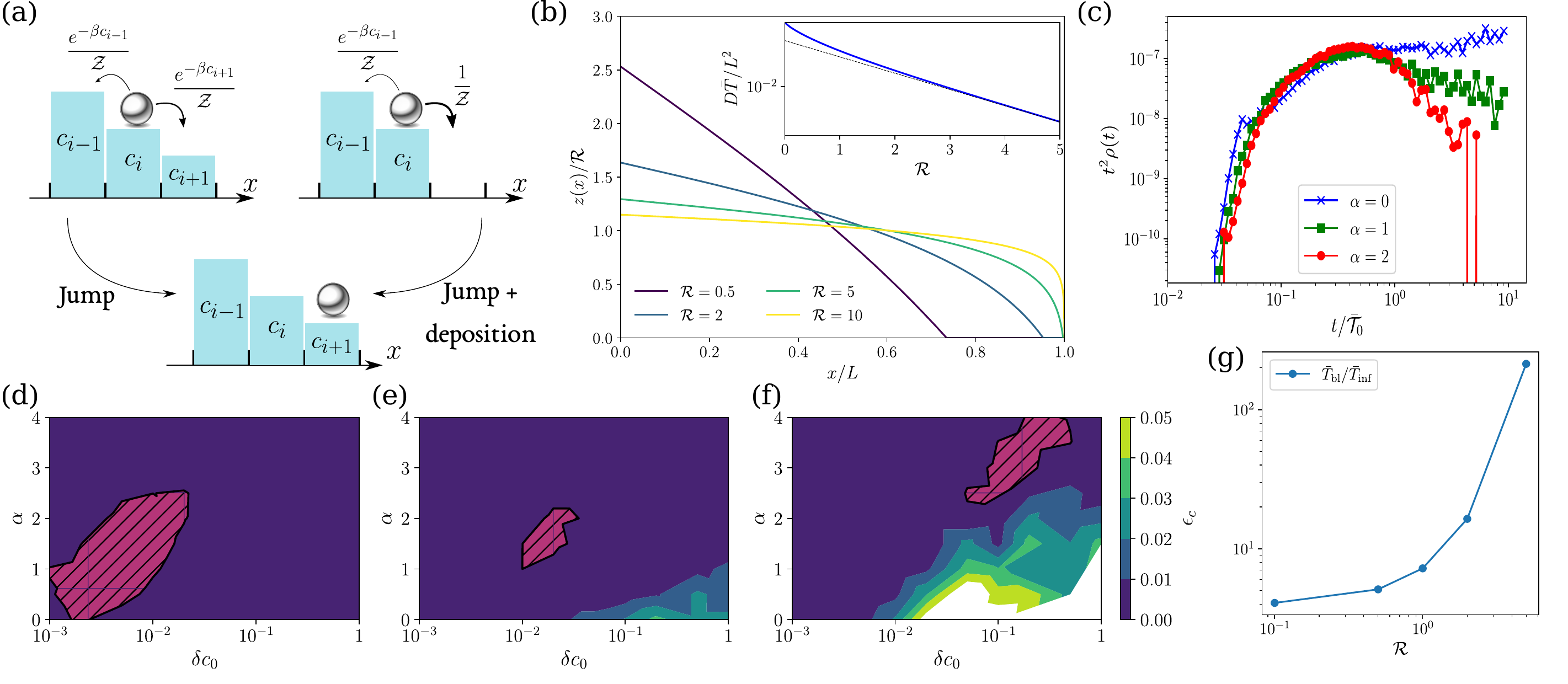}
    \caption{Autochemotactic strategies. (a) Sketch of the model, in the informed protocol. The searcher aims to deposit a predetermined amount on each lattice site. From its current site, it jumps to one of the neighbouring sites $i\pm 1$ with probability $\propto e^{-\beta c_{i\pm 1}}$. Then, if it is the first time this new site is visited, the searcher deposits the amount $c_{i+1}$. If not, it does not deposit anything. (b) Optimal cue profiles in the informed strategies, for various values of the normalized initial supply $\mathcal{R} = \beta R_0$. The larger $\mathcal{R}$ becomes, the flatter the profile is. The inset shows the optimal MFPT (solid line) with its large-$\mathcal{R}$ limit $3\bar{\mathcal{T}}_0e^{-\mathcal{R}/L}/2$. (c) $t^2\rho(t)$, where $\rho(t)$ is the first-passage time distributions in the blind process for $\mathcal{R}=1$, $\beta \delta c_0=0.02$ and various values of $\alpha$. As $\alpha$ grows, the long.time tail becomes thicker and eventually decays slower than $t^{-2}$, making the MFPT diverge. (d)-(f) MFPT in the $(c_0, \alpha)$ plane for $\mathcal{R}=0.1, 2, 5$, in increasing order. The solid line shows where the MFPT is within $5\%$ of its minimum while the color codes for the convergence criterion $\epsilon_c = \bar{T}_c(\tau_c=9\bar{\mathcal{T}}_0)/\bar{T}_c(\tau_c=10\bar{\mathcal{T}}_0)-1$. As $\mathcal{R}$ increases, a larger portion of the parameter space is such that the MFPT diverges. The location of the optimal strategy is indicated by the hatched regions in which the MFPT is within $2\%$ of its minimum sampled value. (g) Ratio$\bar{T}_{\mathrm{bl}}/\bar{T}_{\mathrm{inf}}$ in the optimal strategies as a function of $\mathcal{R}$. Unlike the first two protocols, this ratio grows monotonically as the informed strategy is very efficient while the blind one performs poorly.}
    \label{fig:autochemotaxis}
\end{figure*}

\subsection{Model}

We start by defining the model in discrete space and time. We consider a one-dimensional lattice of size $N_L = L/a$ -- where $a$ is the lattice constant -- over which a concentration field $c$ is defined on each site. The walker has at its disposal a cue supply, whose initial size is $R_0$ and from which it can pick and deposit on lattice sites. The remaining supply at time step $n$ is noted $R_n$. At each time step, the walker first jumps from its current site $i_n$ to one of its two neighboring sites $i_n \pm 1$ with probability 
\begin{equation}
    p_{i_n\to i_{n}\pm 1} = \frac{e^{-\beta i_{n}\pm 1}}{e^{-\beta c_{i_n+1}} + e^{-\beta c_{i_n-1}}}
\end{equation}
For $\beta>0$, the walker favors jumping toward sites with low concentration. After the jump, if $R_n>0$, it can deposit an amount $h_n = a\delta c_n$ on the new site $j_n = i_n \pm 1$ and in turn update the value $c_{j_n}$ of the concentration field by an increment $\delta c_n$ which is subtracted from $R_n$. Otherwise, if $R_n=0$, it does not deposit anything as the cue supply is depleted. $\delta c_n$ is not necessarily constant, it can be random or deterministic, and depend on time and on the current position of the walker. The finite amount of cue in the supply implies that once $R_n$ reaches $0$, i.e. at depletion time, we have
\begin{equation}
    \label{eq:finite_hi}
    \frac{L}{N_L} \sum_{i=0}^{N_L}  c_i = R_0
\end{equation}
 The deposition protocol that determines $\delta c_n$ in conjunction with the constraint (\ref{eq:finite_hi}) impacts significantly the dynamics of the walker and in turn its search performance. \\

In the continuous limit, the partial differential equation that governs the probability distribution $\rho(x,t)$ to find the walker at position $x$ at time $t$, can be shown to reduce to a Fokker-Planck equation with a space- and time-dependent cue field $c(x,t)$ \cite{Meyer2023AlignmentBandFormation}, reading 
\begin{align}
    \label{eq:KS_rho}
    \partial_t \rho(x,t) =& D_p \left[\partial^2_x \rho(x,t) +2\beta \partial_x \left( \rho(x,t) \partial_x c(x,t) \right) \right]
\end{align}
Therefore, $c(x,t)$ effectively acts as a time-dependent potential that is progressively generated by the searcher as it walks. Its time evolution depends strongly on the deposition protocol, as well as on the finite supply constraint, reading
\begin{equation}
    \int_0^L dx c(x,\tau_{\mathrm{dep}}) = R_0
\end{equation}
where $\tau_\mathrm{dep}$ is the depletion time. Note that the effective interaction strength $\beta$ acts as a scaling prefactor to the concentration field, and in turn of the deposition rate $h_n$ and of the cue supply $R_n$. It will therefore be kept fixed to $\beta =1$ throughout the rest of the analysis and will serve as a unit of auto-chemotactic cue. 

\subsection{Informed deposition}

Similarly to the asymmetric flags, we first consider the case where the walker knows where it stands at all times and decides to deposit cue accordingly. For this, we consider the same deposition protocol, which we first describe in the discrete setting. We first decide on a predetermined cue profile $c^*_i$. As a walker reaches site $i$, if it is the first time it reaches it then it deposits an amount $c_i^*$, otherwise it does not deposit anything. Consequently, the walker deposits a non-zero amount only if its current site is the furthest site it has reached so far. A sketch of the protocol is displayed in figure \ref{fig:autochemotaxis}a.

\subsubsection{Deriving the MFPT}
To derive the MFPT for such an informed protocol, we first consider the following, more general process, in which a walker is initialized on the semi-infinite lattice on site $i=1$. We note $M_n$ the farthest site that the walker has reached so far at time $n$. At each time step, the probability of jumping from its current site $i_n$ to any neighboring site $i_n \pm 1$ is given by 
\begin{equation}
    p_{i_n \to i_n \pm 1} = \left\{ 
    \begin{tabular}{l l}
        $q^\pm_{i_n}$ & if $i_n = M_n$ \\
        $\pi^\pm_{i_n}$ & if $i_n < M_n$
    \end{tabular}
    \right.
\end{equation}
where $q^\pm_{i}$ and $\pi^\pm_{i}$ are given space-dependent transition probabilities with $q^-_{i}+q^+_{i}=1$ and $\pi^-_{i}+\pi^+_{i}=1$. Let $\bar{\theta}_M$ be the mean time to go from site $M$ to site $M+1$, given $M$ is the current furthest site reached. It obeys a backward equation of the form
\begin{equation}
    \bar{\theta}_M = q_{M}^{-}\left[1+\bar{\Psi}_M + \bar{\theta}_M \right] + q_M^+ 
\end{equation}
where $\bar{\Psi}_M$ is the mean time to return to site $M$ from site $M-1$. The total mean first passage time $\bar{T}$ to reach the final site $N_L$ is thus found as 
\begin{align} 
    \label{eq:T_vs_theta}
    \bar{T} =& 1+\sum_{M=1}^{N_L-2} \bar{\theta}_M  = 1+\sum_{M=1}^{N_L-2} \frac{1+q_M^- \bar{\Psi}_M }{q_M^+}
\end{align}
where the first term accounts for the reflecting boundary condition at the origin. Because $\bar{\Psi}_M$ corresponds to a process where all sites have already been visited, it only depends on the probabilities $\pi_i^\pm$. It can be inferred by first introducing  the mean time $\bar{\psi}_{j,M}$ to reach site $M$ starting from site $j<M$, such that $\bar{\psi}_{M-1,M} = \bar{\Psi}_M$, which obeys a recursive relation reading
\begin{align}
    \bar{\psi}_{j,M} =& 1 + \pi_{j,M}^+ \bar{\psi}_{j+1,M} +  \pi_j^- \bar{\psi}_{j-1,M} 
\end{align}
Using the boundary condition $\bar{\psi}_{0,M} = \bar{\psi}_{-1,M} + 1$ due to the reflecting wall at the origin, we can solve the recursive relation as
\begin{equation}
    \label{eq:sol_psi_jM}
    \bar{\psi}_{j,M}  = \bar{\psi}_{j-1,M}  - \sum_{k=1}^{j-1} \frac{1}{\pi^+_{k}}\prod_{l=k+1}^{j-1}\frac{\pi^{-}_{l}}{\pi^{+}_{l}} - \prod_{l=1}^{j-1}\frac{\pi^{-}_{l}}{\pi^{+}_{l}} 
\end{equation}
Using $\bar{\psi}_{M,M} = 0$, the above equation for $j=M$ is used to find $\bar{\psi}_{M-1,M} = \bar{\Psi}_M$ and substitute it in Eq.~(\ref{eq:T_vs_theta}) to obtain
\begin{align}    
    \bar{T} =&  1+ \sum_{M=1}^{N_L-2} \frac{1}{q_M^+} \nonumber \\
    &+ \sum_{M=1}^{N_L-2} \frac{q_M^-}{q_M^+}\left[\sum_{k=1}^{M-1} \frac{1}{\pi^+_{k}}\prod_{l=k+1}^{j-1}\frac{\pi^{-}_{l}}{\pi^{+}_{l}} + \prod_{l=1}^{M-1}\frac{\pi^{-}_{l}}{\pi^{+}_{l}}  \right]
\end{align}
In the particular model of cue deposition, the probabilities $q^{\pm}_{l}$ and $\pi^{\pm}_{l}$ are determined in terms of the prescribed concentration field $c^*$, i.e.
\begin{gather}
    q_{i}^- = \frac{e^{-\beta c_{i-1}^*}}{1+e^{-\beta c_{i-1}^*}}, \qquad q_{i}^+ = \frac{1}{1+e^{-\beta c_{i-1}^*}} \\
    \pi_{i}^- = \frac{e^{-\beta c_{i-1}^*}}{e^{-\beta c_{i+1}^*}+e^{-\beta c_{i-1}^*}}, \qquad \pi_{i}^+ = \frac{e^{-\beta c_{i+1}^*}}{e^{-\beta c_{i+1}^*}+e^{-\beta c_{i-1}^*}}
\end{gather}
We therefore obtain, after some algebra
    \begin{align}
        \bar{N} =&  N_L-1+ 2\sum_{M=1}^{N_L-2} \sum_{k=1}^{M} e^{-\beta (c_k^* + c_{k-1}^*)} e^{\beta c_{M}^*}
    \end{align}
In the continuous limit, we have $\bar{T} = \bar{N} dt$ with $dt = a^2/2D$ where $D$ is the diffusion constant kept constant when taking the limit $a, dt\to 0$. This leads to 
\begin{align}
    \label{eq:kramers_autochemo}
    \bar{T}_{\mathrm{inf}} = &  \frac{1}{D}\int_{0}^L dx \int_{0}^x dy   e^{-2\beta c^*(y)} e^{\beta c^*(x)} 
\end{align}
This expression is very similar to Kramers formula for the mean first passage time of a diffusive particle in a potential $c(x)$, i.e. Eq.~(\ref{eq:Kramers}), except for the factor 2 in one of the two exponential factors. This reflects the asymmetry of the process due to the deposition process: the walker experiences the actual potential $c^*(x)$ only if it has already reached $x$ in the past. This additional factor allows for a potential decrease of the MFPT for an appropriately chosen profile $c^*(x)$. \\

\subsubsection{Euler-Lagrange minimization}

We can now ask the following question: what is the best cue profile $c^*(x)$ that minimizes $\bar{T}$? For the sake of clarity, we introduce the normalized field $ z(x) = \beta c^*(x)$ and the Lagrangian $\mathcal{L}$ aimed to account for the finite supply constraint as well as the constraint $z(x)\geq 0$, i.e.
\begin{align}
    \mathcal{L}\left[ z, l, \mu \right] =& \bar{T}_{\mathrm{inf}}\left[ z \right] + l \left( \int_0^L dx z(x) - \mathcal{R}L \right)\nonumber \\
    & + \int_0^L dx \mu(x)z(x) 
\end{align}
where $l$ and $\mu(x)$ are Lagrange multipliers and $\mathcal{R} = \beta R_0/L$. The positivity condition on $z(x)$ implies the Karush-Kuhn-Tucker conditions \cite{Kuhn1951}, namely $\mu(x) z(x) = 0$, $\mu(x) \geq 0$ and $z(x)\geq 0$ for all $x \in [0,L]$. First, we consider a region of space over which we assume $z(x)>0$, and hence $\mu(x) = 0$. The Euler-Lagrange equation reads
\begin{equation}
\label{eq:A_m_2B}
 A(x)  - 2 B(x) + lD = 0
\end{equation}
where we have defined
\begin{align}
    A(x) =&  e^{z(x)} \int_{0}^x dy   e^{-2z(y)}\\
    \label{eq:def_Bx}
    B(x) =& e^{-2z(x)} \int_{x}^L dy   e^{z(y)}
\end{align}
Now, note that
\begin{align}
    \label{eq:zdot}
    z'(x)\left[A(x)+4B(x)\right] =& - 3e^{-z (x)}
\end{align}
which shows that $z(x)$ decreases monotonically over the interval where it is strictly positive. We therefore infer that $z(x)>0$ for $0\leq x< x_\circ$ and $z(x) = 0$ for $x_\circ \leq x \leq L$, where $x_\circ$ remains to be determined. Let us first solve the problem for $0\leq x< x_\circ$. Differentiating Eq.~(\ref{eq:def_Bx}) and using eqs~(\ref{eq:A_m_2B}) and (\ref{eq:zdot}) to substitute $A(x)$ and $e^{-z(x)}$ leads to $B'(x) = -lDz'(x)/3$, which can be integrated to find
\begin{equation}
    B(x) = -\frac{\eta}{3 }\left( z(x) - z_0 \right) + \frac{\eta}{2} 
\end{equation}
where we have used $A(0)=0$ and in turn $B(0) = lD/2$, and defined $z_0= z(0)$. Substituting back $B(x)$ into Eq.~(\ref{eq:zdot}) and using again Eq.~(\ref{eq:A_m_2B}) to eliminate $A(x)$, we obtain
\begin{align}
     z'(x) =& \frac{3 e^{-z(x)}}{2\eta \left[ z(x) - z_0 - 1 \right]  }     
\end{align}
This can be solved with the change of variable $\zeta (x) = z(x) -z_0 - 1$, which obeys
\begin{align}
     \frac{d}{dx}\left[(\zeta (x) - 1) e^{\zeta (x)}\right] =&  \frac{3 e^{-z_0 -1}}{2\eta   }
\end{align}
Using $\zeta(0) = -1$ we obtain
\begin{equation}
    \label{eq:sol_zeta}
    (\zeta (x) - 1) e^{\zeta (x)} =  Hx - 2e^{-1}
\end{equation}
where $H =3 e^{-z_0 -1}/2lD$. Three constants remain to be determined, namely $l$, $z_0$ and $x_\circ$, which can be obtained using the three constraints 
\begin{align}
    \label{eq:const1}
    z(x_\circ) &= 0 \\
    \label{eq:const2}
    \int_0^{x_\circ} z(x) dx &= \mathcal{R}L \\
    \label{eq:const3}
    A(x_\circ) - 2B(x_\circ)  &= -lD 
\end{align}
Eqs~(\ref{eq:const2}) and (\ref{eq:const3}) require to evaluate 2 integrals involving $z(x)$: This is done by using the identity $H = \zeta e^\zeta \frac{d\zeta}{dx}$ that can be formally integrated to find 
\begin{gather}
    \int_{0}^{x_\circ} dx   e^{-2z(x)} = \frac{2lD z_0 }{3}\\
    \int_0^{x_\circ} z(x) dx = \frac{2lD }{3}\left(   (2z_0-3)e^{z_0} +z_0+3 \right) 
\end{gather}
These expressions are inserted into the constraint equations (\ref{eq:const1})-(\ref{eq:const3}) to eventually show that the initial value of the cue profile $z_0$ obeys the equation
\begin{gather}
    \label{eq:z0}
    z_0 - \frac{3}{2}  + \frac{ (2z_0 + 3)^2 e^{-z_0}}{16 - (4z_0 + 10)e^{-z_0}} = \mathcal{R}
\end{gather}
and $l$ and $x_\circ$ follow from it, reading
\begin{align}
     l  =& \frac{6e^{-z_0}}{8 - (2z_0+ 5) e^{-z_0}}\frac{L}{D}\\
     x_\circ =& \frac{4\left[2- (z_0 + 2)e^{-z_0}  \right]}{8 - (2z_0+ 5) e^{-z_0}}L
\end{align}
In the limit $\mathcal{R}\to\infty$, we have $z_0\to\mathcal{R}$, and consequently the concentration profile approaches a rectangular profile over the interval $[0,L]$. This limit can arise either from a very strong interaction or from an abundant supply of cues. In the former case, even a finite amount of cue deposited at the edge of the explored region is sufficient to push the walker towards the next unexplored region, allowing the available cue to be distributed uniformly while maintaining ballistic motion. In the latter case, although the deposition is uniform, the large total supply implies that a substantial amount of cue is deposited at each step. The resulting interaction again strongly drives the walker towards the unexplored region, leading to ballistic propagation.\\

To find the solution for a finite value of $\mathcal{R}$ we can use Eq.~(\ref{eq:sol_zeta}) and transform $\zeta(x)$ back to $z(x)$ to find
\begin{align}
    z (x) =& z_0 + 2 \nonumber \\
    &+ \mathcal{W}_{-1}\left( \left[\left(8 - (2z_0+ 5) e^{-z_0} \right) \frac{x}{4L} - 2\right] e^{-2}\right)
\end{align}
for $x<x_\circ$, where $\mathcal{W}_{-1}$ is the negative branch of the Lambert $\mathcal{W}$ function, and $z(x)=0$ for $x\geq x_\circ$. The corresponding MFPT is then expressed in terms of $z_0$, reading
\begin{equation}
    \bar{T}_{\mathrm{inf}} = \frac{3e^{-z_0}\left(8 - \left(z_0^2 + 5 z_0 + \frac{29}{4}\right)e^{-z_0}\right)}{(8 - (2z_0+5)e^{-z_0})^2} \bar{\mathcal{T}}_0
\end{equation}

In the limit $\mathcal{R}\ll L$ of low cue supply, a series expansion of Eq.~(\ref{eq:z0}) in powers of $z_0$ allows to infer the asymptotic behavior of the quantities at play, namely
\begin{align}
    z_0 =& \sqrt{\frac{3\mathcal{R}}{2}} + O\left(\mathcal{R} \right) \\
    \frac{x_\circ}{L} =& \sqrt{\frac{8\mathcal{R}}{3}} + O\left(\mathcal{R} \right) \\
    \bar{T}_{\mathrm{inf}} =& \bar{\mathcal{T}}_0 \left[ 1 - 2\mathcal{R}\right] +  O\left(\mathcal{R}^{3/2} \right) 
\end{align}
In accordance with the external pulling process and the flag deposition process, we find that a low cue supply require to concentrate the cue deposition in a smaller region as $x_\circ \to 0$ as $\mathcal{R} \to 0$. 

On the other end, in the limit of large supply $\mathcal{R}\gg L$, we find
\begin{align}
    z_0 =& \mathcal{R} + \frac{3}{2} + O\left(\mathcal{R}^2e^{-\mathcal{R}} \right) \\
    \frac{x_\circ}{L} =& 1 - \frac{\mathcal{R}}{4}e^{-\mathcal{R}} \\
    \bar{T}_{\mathrm{inf}} =& \frac{3 \bar{\mathcal{T}}_0}{2}e^{-\mathcal{R}}  +  O\left(\mathcal{R}^2 e^{-2\mathcal{R}} \right) 
\end{align}
The informed strategy is therefore found to be very efficient for this model as the MFPT decays exponentially fast with the size of the cue supply. These results are displayed in figure \ref{fig:autochemotaxis}b, where we show the optimal profiles and the corresponding MFPT.

\subsection{Blind deposition}

Similarly to the flag deposition process, the blind auto-chemotactic process implies that the searcher can deposit cues at positions it has previously visited as soon as $\beta$ is finite. As a result, the effective potential profile $c(x)$ generated by the searcher's trajectory in the pre-depletion stage may be non-monotonic, which may give rise to potential barriers. These can be quickly overcome in the pre-depletion stage as new deposition in a potential valley decreases the height of the barrier. However, in the post-depletion stage, the potential profile remains fixed and the searcher may be trapped by potential barriers and require a significantly long time to escape it. 

We hypothesize that a time-dependent deposition rate could permit to avoid the problem of fixed high potential barriers in the post-depletion stage. In particular, the searcher may adapt the amount it deposits according to its remaining cue supply, and we therefore choose a deposition scheme of the form 
\begin{equation}
    \delta c_n = \delta c_0 \left(\frac{R_n}{R_0} \right)^\alpha
\end{equation}
where $\delta c_n$ is the increment in the cue profile deposited at the searcher's site at time step $n$, in the discrete setting, which is then subtracted from the cue supply. 
Three regimes can be distinguished: (i) for $\alpha< 1$, depletion occurs at a finite time $\tau_\mathrm{dep} = R_0/L\delta c_0(1-\alpha)$; (ii) for $\alpha=1$ the cue supply decreases exponentially at a rate $R_0/h_0$ such that depletion is technically never reached although the amount deposited become increasingly negligible at long times; (iii) for $\alpha>1$, the cue supply decreases algebraically with an exponent $1/(1-\alpha)$, again without ever reaching full depletion. The time dependence of the deposition rate is expected to directly influence the cue profile generated by the walk. As $\alpha$ increases, the deposition rate decays more rapidly, leading to initially steeper profiles that may benefit the search. However, this potential benefit is counteracted by two competing effects. First, increasing $\alpha$ does not entirely prevent the formation of barriers induced by trajectory fluctuations. Second, the rapid decay of the deposition rate means that cue deposition becomes negligible relatively early, leaving a substantial fraction of the process with little further contribution from the resource.\\

To quantify these effects, we run simulations of the process on a lattice of $N_L = 1000$ sites, and sample various values of $\alpha$ and $\delta c_0$ for given values of the initial cue supply $R_0$. In a substantial portion of the parameter space, the long-time tail of the first-passage time distribution $\rho(t)$ decays algebraically with $t$ with an exponent larger than $-2$, as shown in figure \ref{fig:autochemotaxis}c. In these cases, the first moment of the distribution diverges and the MFPT is thus undefined. This occurs for parameters where the cue profile at the time of depletion is likely to contain high barriers such that post-depletion self-trapping leads to very large search times and quantitatively impact the long-time trails of the distribution. Numerically, we discriminate between  converging and diverging MFPT by sampling the cutoff-restricted mean, i.e. we compute the mean first passage time $\bar{T}_c$ of subsets of trajectories whose search time is less than a cutoff $\tau_{c}$. We then test whether $\bar{T}_c$ converges as $\tau_c$ increases. In practice, we consider that the MFPT is convergent if $\bar{T}_c$ increases by less than $0.1\%$ between $\tau_c = 9\bar{\mathcal{T}}_0$ to $\tau_c = 10\bar{\mathcal{T}}_0$. This is of course an arbitrary criterion but it captures well the transition between converging and diverging MFPTs, as shown in figures \ref{fig:autochemotaxis}d-f. 

The optimal strategies that we identify are such that both $\delta c_0$ and $\alpha$ increase with $R_0$, although the precise location of the minimum is difficult to locate numerically due to finite sampling and a slow variability of the MFT around its minimum in the $(\delta c_0, \alpha)$-plane, see figure \ref{fig:autochemotaxis}d-f. This implies that for the blind strategies, the larger the supply is, the more one should deposit in the early stages of the process, relative to the initial supply. This strategy is precisely opposite to the optimal informed one, where the optimal cue profile becomes flatter as the size of the supply increases. Such an optimal strategy is entirely aimed at avoiding high potential barriers in the post-depletion stage. As the initial supply is larger and hence the potential deposition rate grows, the risk of creating large barriers becomes so important that depositing much cue at short times allows to limit the probability of generating them. This strategy is purely a risk-averse one as the possibility of being self-trapped is significantly more damageable than the gain obtained by speeding up the fastest trajectories. The decrease of the optimal MFPT as a function of the initial supply $R_0$ is therefore slow. In comparison, the informed strategy is orders of magnitude more efficient and the ratio $\bar{T}_{\mathrm{bl}}/\bar{T}_{\mathrm{inf}}$ for the optimal strategies grows monotonically with $R_0$ and diverges as $R_0 \to\infty$, as shown in figure \ref{fig:autochemotaxis}g.

\section{Discussion}

\subsection{Random target}
The results presented  so far hold for a fixed target position. However, the term search traditionally refers to a process in which the target may be randomly placed according to a particular probability distribution $\psi_T(x_T)$. Considering a target homogeneously distributed across a fixed interval $[0,L]$, i.e. $\psi_T(x_T) = L^{-1}\Theta(L-x_T)$, the phenomenology of the results obtained for the fixed-target strategies is preserved. Details of calculations and simulations are provided in the Supplemental Material \footnote{The Supplemental Material provides details of the derivations and numerical analysis of the optimization problem when considering a randomly placed target}, that we summarize below.

\subsubsection{External pulling}
For the external pulling process, the informed strategy can be obtained numerically using the same technique as for the fixed target. The MFPT can be written in terms of a discretized potential with adjustable locations $x_k$ for the potential jumps, resulting in a cubic polynomial in $x_k$. The optimal MFPT can therefore not be found by means of a matrix inversion problem as previously, but only using numerical optimization. We find however, that the resulting optimal MFPT is very close to the MFPT obtained by optimizing a 3-part piecewise potential containing a constant plateau $V(x) = W$ up to $x=x_-$, a linear ramp from $W$ to 0, between $x_-$ and $x_+$ and a final constant part at zero potential from $x_+$. In the limit $W\to 0$, both $x_+$ and $x_-$ tend to $L/3$ with a corresponding MFPT $\bar{T} = \bar{\mathcal{T}}_0/3$. As $W \to \infty$, both $x_-$ and $L-x_+$ decay as $(\beta W)^{-1}$, similarly to the fixed-target scenario. 

For the blind strategy, the problem can be solved analytically by setting the switching time $\tau = 0$. The resulting MFPT is minimized for an optimal force $F$ that grows as $F = W/L+ 1/2\beta L + O((\beta W)^{-1})$ as $W \to \infty$, yielding a mean search time that decays $\bar{T} = \bar{\mathcal{T}}_0 /\beta W$.

\subsubsection{Directional flags}
As for the fixed-target case, the informed strategy is straightforwardly inferred from the derivation of the optimal pulling protocol presented in section \ref{sec:informed_pulling}. Our conclusions are therefore the same for the search problem, and we obtain that the optimal deposition protocol is shifted to the left and leads to a MFPT that decays inversely proportionally to the number of flags. The blind search is also found to be optimized for faster deposition protocols than for the exit problem, which still scales as $N^{-2}$ in the large $N$ limit and results in an optimal search time that also decays as $N^{-1}$. As for the fixed-target case, the ratio between the optimal blind and informed MFPT increases with the number of flags and reaches a plateau as $N\to\infty$. 

\subsubsection{Auto-chemotaxis}
For the auto-chemotactic process, the conclusions for the fixed-target case still hold. In the informed strategy, the optimal cue profile cannot be found analytically but only by means of numerical optimization. We still find that it is strictly positive for values of $x$ lower than a certain $R_0$-dependent threshold, and $0$ beyond it, as for the fixed-target case. However, this threshold is lower than for the fixed-target case, as expected. The profile becomes flatter as $R_0$ grows, similarly to the fixed-target case, and the resulting MFPT decays exponentially with $R_0$.  In the blind strategies, the problem of self-trapping still hinders the search performance such that the optimal MFPT decreases with $R_0$ but much more slowly than for the informed process.

\subsection{Dimensionality}

As for many stochastic processes, the dimensionality of the system may impact its dynamical properties. For first-passage problems, a crucial change is the following: in dimension 1, a walker initialized at position $x_0$ needs to go through all positions between $x_0$ and $x_f$ in order to first reach position $x_f$. This makes it possible to derive backward recursive equations for first-passage properties. This is no longer the case in higher dimensions, as a walker has an infinite number of possible paths to go from one position to another. This being clarified, we can still make a few comments and conjectures about our particular resource-usage problem.

The external pulling protocol can be generalized to higher dimensions by applying a potential whose minimum is located at the target and whose extremal values equal the work budget. However, analytical expressions may be more difficult to obtain, as an exact, compact Kramers-like formula is no longer available. Still, an optimal potential should exist, for both the informed and blind protocols. In the informed case, the piecewise form of the potential may survive, as its efficiency resides mainly in the steep potential descent it creates when little resource is available. For the blind process, however, results might change. In dimension 1, the optimal protocol is such that a force directed towards the target is switched on almost immediately, and its amplitude is adjusted to exhaust the work budget exactly when the target is reached. In higher dimensions, this strategy may be reproducible only in certain settings, namely if the search domain is circular and the searcher is initialized at its border. Otherwise, the position the searcher reaches when the budget is exhausted may not be the target location.

The model of directional flags is more complicated to generalize to higher dimensions. In dimension 1, a flag is a point-like object against which the searcher is reflected. In higher dimensions, two natural generalizations can be designed. First, flags could be defined as reflecting lines circling around the target. Although this generalization is mathematically sound and its resulting physics may be close to that of the original one-dimensional model, it is in practice far from any realistic natural process. The second option is to still define flags as point-like objects and assign them an influence radius, such that the walker is redirected towards the target if it enters the vicinity of the flag. Unless the influence radius is sufficiently large, we expect such flags to be substantially less efficient than in the one-dimensional case, as the searcher can travel around them. In addition, the analytical treatment may be much more difficult, as the problem cannot be mapped onto diffusion in a potential: a force field that is constant in a disk and zero otherwise is, for instance, not conservative. We can therefore not easily generalize our results to higher dimensions for the case of directional flags.

On the contrary, auto-chemotactic cues can easily be generalized to higher dimensions, and have been for many years. The true self-avoiding walk has been investigated on many occasions in dimensions 2 and 3, and its dynamical properties are well established \cite{Amit1983, Grassberger2017}. The Keller-Segel equation, i.e. the standard continuous model for chemotaxis, has also been studied extensively over the past decades \cite{Keller1970,Arumugam2020}. The self-trapping mechanism observed in the blind strategies in dimension 1 remains in higher dimensions but becomes weaker as the searcher has more ways to escape. Generalizing our resource-constrained search problem to higher dimensions would affect the informed strategy, as temporary self-trapping could still occur while the cue is being deposited to reproduce an optimal profile. This is not the case in dimension 1, which makes the strategy extremely efficient there. The blind strategies would not be majorly affected, as the detrimental post-depletion self-trapping would also occur. It is, however, not clear how the long-time tails of the first-passage time distribution would be altered, and under which conditions the MFPT would be finite or diverge.

\subsection{Conclusion}

The results presented in this work highlight that the optimal use of finite resources depends on both the nature of the resource and the information available to the searcher. informed strategies are found to require resource allocation to be concentrated in regions of space whose size become vanishingly small as the supply size decreases. Conversely, blind strategies are optimized when resource consumption is evenly spread in time and enhance the likelihood for the supply to be exhausted close to target arrival. As expected, informed strategies, in which the searcher knows its position at all times and can adapt resource deposition accordingly, generally outperform blind strategies. However, the magnitude of this advantage depends strongly on the type of resource being used. 

For both directional flags and auto-chemotactic walks, the key limitation of blind strategies is that not only the location of deposited cues may fluctuate from one trajectory to an other, but their spatial ordering no longer necessarily reflects the order in which they were deposited. The searcher may therefore expend resources at locations it has already visited, rather than using them to improve its future search. The consequences of this inefficiency, however, differ between the two mechanisms. Directional flags remain beneficial because they consistently bias the searcher towards the target, even when deposited at previously visited locations. By contrast, auto-chemotactic cues can generate complex potential landscapes in which poorly placed deposits hinder subsequent motion and lead to self-trapping.

More broadly, our findings highlight the importance of jointly considering resource dynamics, environmental memory, and the searcher’s access to information when designing efficient search strategies with finite resources.

\section{Acknowledgements}
This work was supported by the Medical Research Council (Grant No. MR/Y003845/1) and the Engineering and Physical Sciences Research Council (Grant No. EP/ V031201/1)


%

\clearpage

\onecolumngrid

\begin{center}
	{\bf \Large Efficient Search with Finite Resources: Optimal Informed and Blind Strategies \\ \vspace{1ex}
		\large Supplemental Material}\\
	
	\vspace{2ex}
	
	\normalsize 
	Hugues Meyer \\
	
	{\it \footnotesize School of Physics and Astronomy, University of Nottingham, Nottingham NG7 2RD, United Kingdom \\ Centre for the Mathematics and Theoretical Physics of Quantum Non-Equilibrium Systems, University of Nottingham, Nottingham NG7 2RD, United Kingdom}
\end{center}
\vspace{5ex}

In this additional section, we revisit the results presented in the main text in processes where the target is randomly distributed across a search domain. The searcher is still initialized at $x=0$, but the target can be localized at any point between $x=0$ and $x=L$ with equal probability. Given the MFPT $\bar{T}(l)$ for a fixed target at position $x=l$, the target-averaged MFPT reads
\begin{equation}
\bar{\bar{T}} = \frac{1}{L} \int_0^L dl \bar{T}(l)
\end{equation}
In most cases, this introduces nonlinearities in the derivations of optimal profiles or time protocols. We therefore rely on numerical optimization in these cases. 

\section{External pulling}

\subsection{Informed strategy}
\begin{figure}
	\centering
	\includegraphics[width=\linewidth]{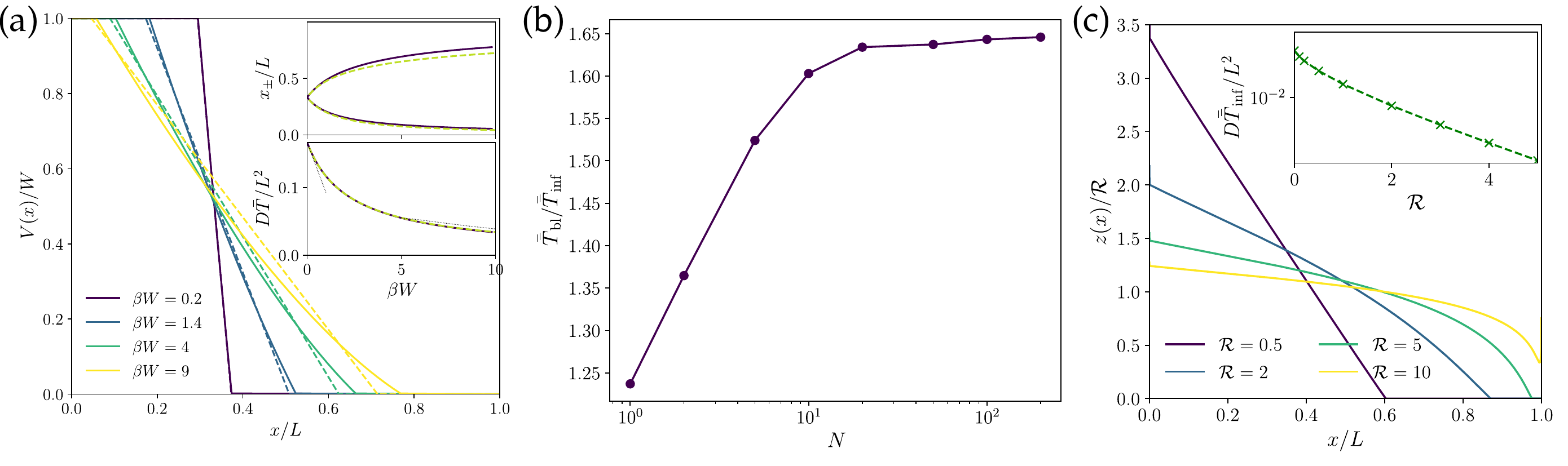}
	\caption{(a) Optimal force profiles for a randomly located target in the informed pulling protocol for various values of the work budget. The solid lines show the result obtained with functional optimization while the dotted lines are the optimal piecewise linear profile. The top inset shows the dependence of $x_\pm$ as a function of the work budget in both cases, while the bottom inset shows the optimal mean search time. (b) Ratio $\bar{\bar{T}}_{\mathrm{bl}}/\bar{\bar{T}}_{\mathrm{inf}}$ in the directional flag process for $p=1$ for a randomly located target, as a function of the number of flags. It increases and reaches a plateau value at large $N$. (c) Optimal cue profiles in the informed auto-chemotactic process with a randomly located target, for various values of the cue budget $\mathcal{R}$. The inset shows that the optimal MFPT decays exponentially with $\mathcal{R}$.}
	\label{fig:SM}
\end{figure}

For the external pulling scenario, the MFPT can be minimized by using the same technique as for the fixed-target case. We introduce a piecewise potential $V^{(N)}(x; x_1, \cdots, x_{N}) = \frac{W}{N} \sum_{k=1}^{N} \Theta\left(x_k - x \right)$ where $\Theta$ is the Heaviside function, and the corresponding target-averaged MFPT reads

\begin{align}
\label{eq:piecewise_potential_search}
\bar{T}_{\mathrm{inf}}  =&  \sum_{k=0}^N  \left[\sum_{l=0}^{k-1}  \rho^{k-l}(x_{l+1}-x_l)\left(x_{k+1}-x_k - \frac{x_{k+1}^2-x_k^2}{2L} \right) + \frac{(x_{k+1}-x_k)^2(3L - 2x_{k+1} - x_k)}{6L} \right] 
\end{align}
where $\rho = e^{-\beta W/N}$. This expression is cubic in $x_k$ and can therefore only be minimized numerically. Similarly to the fixed-target case, we also obtain in the limit $N\to\infty$ as an optimal solution a piecewise potential that equals $W$ for $x<x_-$ and $0$ for $x>x_+$. For   $x_- \leq x\leq x_+$, the potential is not exactly linear but slightly convex, as shown in fig \ref{fig:SM}. We can estimate the locations of $x_\pm$ by approximating the optimal potential by a piecewise linear potential that interpolates between $W$ and $0$ in the interval $[x_-, x_+]$, and obtain 
\begin{align}
\bar{\bar{T}}_{\mathrm{inf}} =&\left[ \frac{(L-x_+)^2 }{2wLD} - \frac{(x_+-x_-)(L-x_+)}{w^2LD}  + \frac{(x+-x_-)^2 }{w^3LD} \right] ((1+w)x_- - x_+)e^{-w} \nonumber\\
&+ \frac{(L-x_+)^3 + 3Lx_-^2 - 2x_-^3}{6LD}+ \frac{(L-x_-)(L+x_-)(x_+-x_-)}{2wLD} -  \frac{(x_+-x_-)^2}{w^2D} + \frac{(x_+-x_-)^3}{w^3LD}
\end{align}
where we recall $w=\beta W$. The solution of the coupled equations $\partial_{x_-} \bar{\bar{T}} = 0$ and $\partial_{x_+} \bar{\bar{T}} = 0$ can be approximated in the limits $w\to 0 $ and $ w\to \infty$. We find for $w\ll 1$ 
\begin{equation}
x_{\pm} = \frac{L}{3} \left[1 \pm \frac{2w}{3} + O(w^2) \right], \qquad \bar{\bar{T}}_{\mathrm{inf}} = \frac{L^2}{D} \left[\frac{1}{6} - \frac{2w}{27} + O(w^2) \right]
\end{equation}
and for $w\gg 1$
\begin{equation}
x_{-} = \frac{L}{2w} + O(w^{-2}), \qquad L - x_+ = \frac{L}{\sqrt{w}}  + O(w^{-1}), \qquad \bar{\bar{T}}_{\mathrm{inf}} = \frac{L^2}{2Dw} + O(w^{-3/2}) 
\end{equation}
The profiles and corresponding MFPT are shown in figure \ref{fig:SM}a.

\subsection{Blind strategy}
For the blind protocol, because the optimal strategy for the fixed-target case is nearly optimal as one switches on the force $F$ instantaneously from the start of the process regardless of the distance from the target, we investigate only this case for the randomly located target. We therefore simply need to average the piecewise function of $F$ obtained for the fixed-target case over all possible target locations to find
\begin{equation}
\bar{\bar{T}} =\frac{L^2}{f^3D} \times\begin{cases}
1-f+\frac{f^2}{2}-e^{-f}  & \text{if } f<w\\
1 - \frac{w^2}{2} + \frac{w^3}{3}  - (w+1)e^{-w}  - \left(1-w+\frac{w^2}{2} - e^{-w}\right)f  + \frac{f^3}{6} & \text{if } f>w
\end{cases}
\end{equation}
where $f=\beta F L$. The minimum is reached for 
\begin{equation}
f =   \frac{6 - 3w^2 + 2w^3 -6(w+1)e^{-w} }{  4-4w+2w^2-4e^{-w}} > w
\end{equation}
For $w\ll 1$ we can expand this result and the corresponding optimal MFPT to find
\begin{equation}
f = \frac{9w}{8} + O(w^2), \qquad \bar{\bar{T}} = \frac{L^2}{D} \left[\frac{1}{6} - \frac{32}{729} w + O(w^2) \right]
\end{equation}
For $w\gg1$ we have
\begin{equation}
f = w + \frac{1}{2} + O(w^{-1}), \qquad \bar{\bar{T}} = \frac{L^2}{2Dw}  + O(w^{-2})
\end{equation}
Similarly to the fixed-target case, the blind and informed optimal MFPT decay as $L^2/2Dw$ for large work budgets and their ratio therefore tends to $1$.

\section{Directional flags}

\subsection{Informed strategy}
Equation~(\ref{eq:piecewise_potential_search}) is valid for the directional flags if we use $\rho = 1-p$, where $p$ is the probability to be reflected against a flag, as described in the main text. Considering a flag profile of the form $x_k = x_- + (k-1)\delta$, i.e. the first flag is deposited at position $x = x_-$ and the next ones are deposited at constant intervals $\delta$ as in the fixed-target case, we can evaluate equation~(\ref{eq:piecewise_potential_search}) in terms of $x_-$ and $\delta$. The resulting expression is very large and is again cubic in $x_-$ and $\delta$. However, we can first investigate the case $N=1$ where we find that $x_-=L/3$ is optimal. Then, for $N\gg 1$, we find, optimally
\begin{equation}
x_- = \frac{L}{2N} + O(N^{-2}), \qquad \delta =  \frac{L}{N} + O(N^{-2}), \qquad \bar{\bar{T}} = \frac{L^2(2-p)}{4DNp} + O(N^{-2})
\end{equation}

\subsection{Blind strategy}

Similarly to the fixed-target case, the blind strategies cannot be investigated analytically in general. We therefore rely on Brownian dynamics simulations to identify optimal strategies. As for the fixed-target case, we find that protocols where the deposition time interval is constant are nearly optimal. The MFPT then decays as $1/N^{-2}$ as $N\gg 1$ but not as quickly as for the informed strategies. In fact, the ratio $\bar{\bar{T}}_{\mathrm{bl}}/\bar{\bar{T}}_{\mathrm{inf}}$ grows monotonically with $N$ and reaches a plateau value as $N\to \infty$, as shown in figure \ref{fig:SM}b.

\section{Auto-chemotaxis}

\subsection{Informed strategy}

In the informed strategies for the randomly located target, we can write the MFPT as
\begin{equation}
\bar{\bar{T}} = \frac{1}{L} \int_0^L dx \int_0^x dy e^{\beta(c(x)-2c(y))}\left(1 - \frac{x}{L}\right)
\end{equation}
where $c(x)$ is the cue profile that one aims to reproduce as the searcher explores further regions of space. The Euler-Lagrange equation with the constraints $c(x)\geq 0$ and $\int_0^L dx c(x) = R_0$ can however not be solved analytically, in contrast to the fixed-target case. We therefore rely on numerical optimization using gradient descent optimization with constraints. We then find that the optimal profiles follow the same phenomenology as for the fixed-target case, namely $c(x)>0$ for $x<x_\circ$ and $c(x)=0$ for $x\geq x_\circ$, where $x_\circ$ is a threshold position that grows with $R_0$ but is smaller than for the fixed-target case. Also, the profile becomes flatter as $R_0$ increases and tends to a constant profile $c(x) = R_0/L$ in the limit $R_0\to\infty$. The corresponding optimal MFPT decays exponentially with $R_0$, as for the fixed-target case. Details of the profiles and corresponding MFPTs are shown in figure \ref{fig:SM}c.

\subsection{Blind strategy}
The blind strategy suffers from the same self-trapping problem as for the fixed-target case. The cue profiles generated by the searcher's walk may present barriers and optimal strategies aim to avoid this. As a result, the optimal MFPT decreases only mildly with $R_0$ and the ratio between the optimal blind and informed strategies grows monotonically with $R_0$, similarly to the fixed-target case.

\end{document}